\documentclass[twocolumn,aps,pra,superscriptaddress]{revtex4-1}
\usepackage{graphicx}
\usepackage{amsmath}
\usepackage{amssymb}
\usepackage{braket}
\usepackage{bm}
\usepackage[colorlinks=true,linkcolor=blue,citecolor=blue,urlcolor=blue]{hyperref}

\begin{document}

\title{Tight-Binding Kondo Model and Spin-Exchange Collision Rate of 
Alkaline-Earth Atoms in a Mixed-Dimensional Optical Lattice}

\author{Ren Zhang}
\email{renzhang@xjtu.edu.cn}
\affiliation{School of Science, Xi'an Jiaotong University, Xi'an 710049, China}
\affiliation{Shaanxi Province Key Laboratory of Quantum Information and Quantum Optoelectronic Devices, Xi'an 710049, China}

\author{Peng Zhang}
\email{pengzhang@ruc.edu.cn}
\affiliation{Department of Physics, Renmin University of China, Beijing, 100872, China} 
\affiliation{Beijing Computational Science Research Center, Beijing, 100084, China}

\date{\today}

\begin{abstract}
We study the two-body problem of ultracold fermionic alkaline-earth (like) atoms in the electronic $^1$S$_0$ state ($g$-state) and $^3$P$_0$ state ($e$-state) which are confined in a quasi-one-dimensional (quasi-1D) tube simultaneously, where in the axial direction the $g$-atom experiences a 1D optical lattice and the $e$-atom is localized by a harmonic potential. Due to the nuclear-spin exchange interaction between the $g$- and $e$-atom, one can use such a quasi-(1+0)D system to realize the Kondo effect. We suggest two tight-binding models for this system, for the cases that the odd-wave scattering between the $g$- and $e$-atom is negligible or not, respectively. 
Moreover, we give a microscopic derivation for the inter-atomic interaction parameters of these models, by explicitly calculating the quasi-(1+0)D low-energy scattering amplitude of the $g$- and $e$-atom in this system and matching this exact result with the ones given by tight-binding models. 
 We illustrate our results for the experimental systems of ultracold $^{173}$Yb and $^{171}$Yb atoms and show the control effect of the confinement potentials
on these model parameters. 
Furthermore, the validity of the simple ``projection approximation" is examined. In this approximation, one derives the interaction parameters of the tight-binding models by directly projecting the 3D Huang-Yang pseudopotential on the ground state of the confinement and the lowest band of the optical lattice. This approximation is supposed to be correct when the 3D inter-atomic scattering length $a_s$ is much smaller than the characteristic lengths (CLs) of the confinements. However, we find that for our system this approximation already does not work when $a_s$ is only of the order of 10\% of the confinement CLs. Using the exact two-atom scattering amplitude, we also calculate the spin-exchanging rate (i.e., the cross-section of the spin-exchanging collision between the $g$- and $e$-atom) for the recent experiment (L. Riegger, {\it et. al.,} Phys. Rev. Lett. {\bf 120}, 143601 (2018)) of $^{173}$Yb atoms in this quasi-(1+0)D system, and investigate finite-quasi-momentum effect of the $g$-atoms in this experiment. Our results show that this effect is very significant.
\end{abstract}

\maketitle

\section{Introduction}

\begin{figure}
\centering
\includegraphics[width=0.4\textwidth]{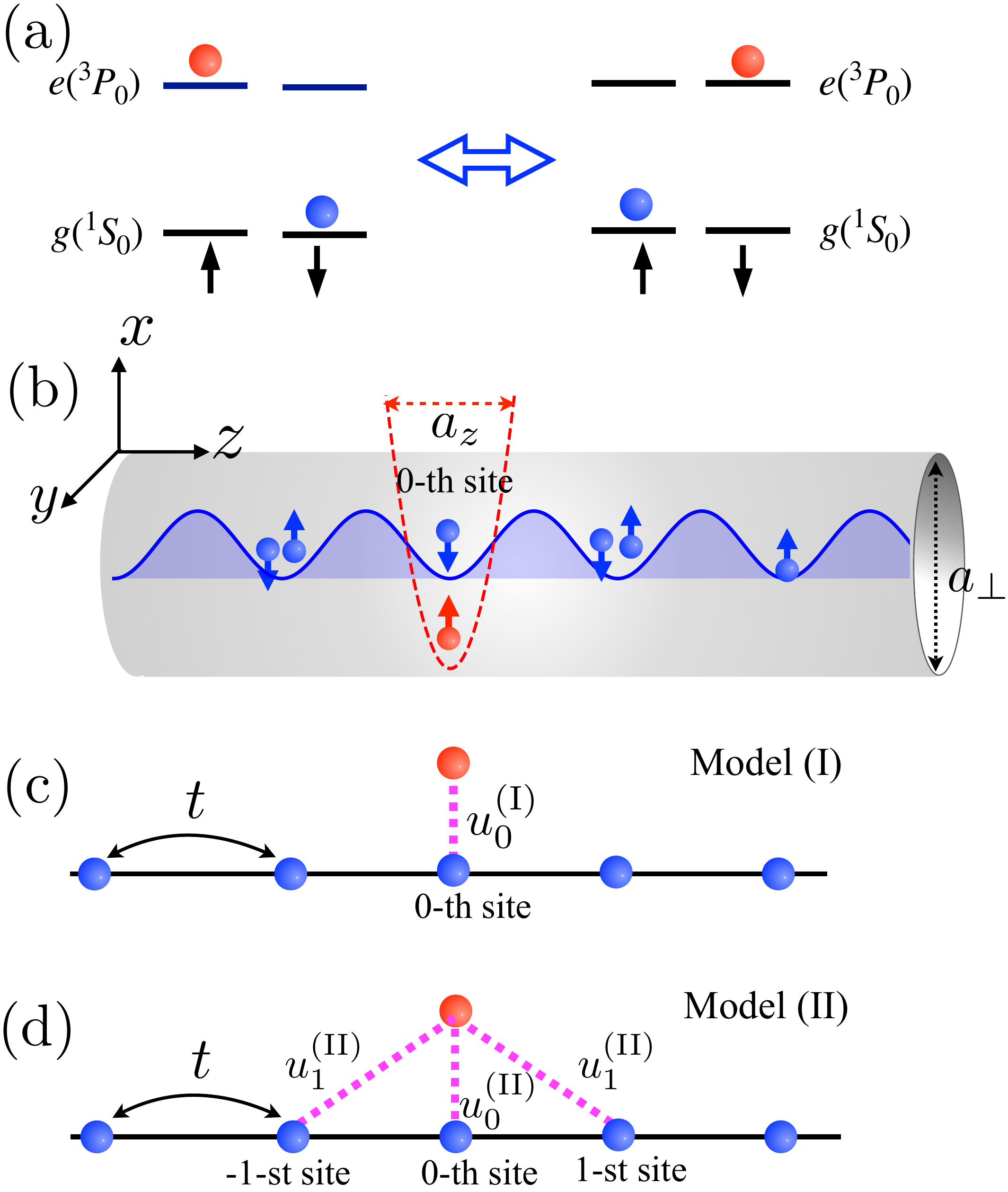} 
\caption{(color online) {\bf (a):} Schematic illustration of spin-exchange interaction between alkaline-earth atoms in $^1{\rm S}_0$ ($g$) and $^3{\rm P}_0$ ($e$) states. The nuclear-spin states are denoted as $\uparrow$ and $\downarrow$ {\bf (b):} The quasi-(1+0)D system studied in this work.
The $g$-atoms (blue balls) and the single $e$-atom (red balls),
which can be in either the nuclear-spin states $\uparrow$ or $\downarrow$,
 are simultaneously confined in the quasi-1D tube. Along the axial direction ($z$-direction), 
the $g$-atoms experience a 1D optical lattice, while the $e$-atom is localized by a harmonic trap centered at the 0-th site of this lattice.
{\bf (c)} and {\bf (d):} Two 1D tight-binding models we suggest for our system.
The $g$-atom hops between the nearest blue sites, with hopping rate $t$. The $e$-atom, which is localized in the $0$-th site, is represented as the red ball.
 {\bf (c) :} In model (I) the interaction between the $e$- and $g$-atoms occurs only when the $g$-atom is at the 0-th site, with interaction strength $u_0^{\rm (I)}$. This model can be used for the cases that the inter-atomic odd-wave scattering can be neglected. {\bf (d) :} In model (II) 
the $e$-atom can interact with the $g$-atom at either the 0-th site or the $\pm 1$-st sites, with with interaction strengths $u_0^{\rm (II)}$ and $u_1^{\rm (II)}$, respectively. This model can reproduce both the even-wave and the odd-wave inter-atomic scattering amplitudes. \label{schematic}}
\end{figure}

Recently, a series of theoretical and experimental works demonstrated that the ultracold gases of fermionic alkaline-earth-(like) atoms are good platforms for the quantum simulation of many-body models with spin-exchange interaction, e.g., the Kondo model \cite{renkondo1,yantingkondo,renkondo2,kondo_exp,Yb173as,Yb171,spin-exchange1,spin-exchange2,XiboSUN,ReyNP,Kuzmenko}. These atoms can be prepared in either the electronic ground state ($^{1}$S$_{0}$ state) or the long-lived electronic excited state ($^{3}$P$_{0}$ state). In addition, there exists a nuclear-spin exchange interaction between two atoms in these two electronic states (Fig.~\ref{schematic}(a)). The experimental measurements show that the spin-exchange interactions of $^{87}$Sr, $^{173}$Yb and $^{171}$Yb atoms are pretty strong, with corresponding three-dimensional (3D) scattering lengths being as large as $10^2a_0$-$10^3a_0$ with $a_0$ being the Bohr's radius \cite{Yb173as,Yb171,spin-exchange1,spin-exchange2,XiboSUN}. In 3D free space
the spin-exchange interaction of $^{87}$Sr and $^{173}$Yb atoms are ferromagnetic \cite{Yb173as,spin-exchange1,spin-exchange2,XiboSUN}, while the one of $^{171}$Yb atoms is antiferromagnetic \cite{Yb171}. Furthermore, with the help of confinement-induced resonance (CIR) \cite{Olshanii1998,Bergeman2003,Haller2010}, one can control the spin-exchange interaction 
by tuning the trapping potentials \cite{renkondo1,yantingkondo,renkondo2,kondo_exp}. This technique has already been realized in the recent experiment of ultracold $^{173}$Yb atoms \cite{kondo_exp}.

In our previous works \cite{renkondo1,yantingkondo,renkondo2} we studied the control of spin-exchange interaction for alkaline-earth-(like) atoms in a mixed-dimensional system via the CIRs. In these works, we assume the atoms are confined in a 2D harmonic potential (i.e., a quasi-1D tube), which has the same frequency for atoms in both $^{1}$S$_{0}$ ($g$-) and $^{3}$P$_{0}$ ($e$-) states. Besides, in the axial direction the $e$-atom further experiences a harmonic trapping potential while the $g$-atom is freely moving. This system corresponds to a (1+0)D effective model where the $g$-atoms are moving in the 1D {\it continuous} space and the $e$-atom is a fixed pointe-like magnetic impurity, with spin-exchange interaction between the $g$- and $e$-atoms in both even-wave and odd-wave channels. It has been shown that this model can be used for the studies of various phenomena based on the Kondo model \cite{kondobook,Kondosolution,Kondosolution1,Kondosolution2}.

On the other hand, the tight-binding {\it lattice models} are also very important for the research of the Kondo effect. For instance, these models are believed to describe the physics in the heavy-fermion and related system \cite{kondobook,Kondolattice,KMLreview}. For ultracold alkaline-earth-(like) atoms, one can realize the 1D tight-binding lattice model with Kondo effect via the mixed-dimensional setup described above, by adding a 1D optical lattice potential for the $g$-atom in the axial direction of the tube (Fig.~\ref{schematic}(b)). This system also has been realized in the experiment of Ref. \cite{kondo_exp}.


In this work, we present the appropriate tight-binding models corresponding to the above mixed-dimensional lattice system (Fig.~\ref{schematic}(b)). In this system, there are both even-wave and odd-wave scatterings between the $g$-and $e$-atoms. Nevertheless, the odd-wave scattering can usually be neglected if the system is not under an odd-wave resonance. The tight-binding model (model I) for this case can only include the interaction between the $e$-atom and the $g$-atom in the same site (Fig.~\ref{schematic}(c)). On the other hand, if the odd-wave scattering is not negligible, in the corresponding tight-binding model (model II) the $e$-atom interacts with the $g$-atom which is either in the same site or in the nearest neighbor sites (Fig.~\ref{schematic}(d)), so that both the even-wave and the odd-wave scattering amplitudes can be reproduced by this model.

Furthermore, we give a microscopic derivation for the interaction parameters of the above two tight-binding models. To this end, we construct the Green's function for our quasi-(1+0)D lattice system, in aid of the 
Mathieu function. Using this Green's function we calculate the exact low-energy scattering amplitudes between the $g$- and $e$-atoms by solving the Lippman-Schwinger equation. In particular, in our calculation the virtual transitions to the excited 
states of the transverse confinement, the excited bands of the axial optical lattice for the $g$-atom and the excited states of the axial trapping for the $e$-atom are explicitly taken into account.
By equalizing these exact scattering amplitudes and the ones given by the tight-binding model, we obtain the interaction parameters. Our results show that as in other quasi-low-dimensional systems, these interaction parameters can be efficiently controlled by the trapping potentials and the lattice potential. For ultracold $^{171}$Yb and $^{173}$Yb atoms we illustrate the interaction parameters for typical experimental cases.

Moreover, when the scattering length $a_s$ between the $g$- and $e$-atoms in the 3D free space is very small, we can approximately derive the interaction parameters by projecting the Huang-Yang pseudopotential in the ground states of the trapping potential and the lowest band of the optical lattice. In this work, we further investigate the applicability of this widely-used ``projection approximation" for our system, and find that this approximation already fails when $a_s$ is only of the order of 10\% of the characteristic lengths of the trapping/lattice potentials, and thus cannot be used for many realistic cases.

In addition, we also use the exact scattering amplitude to study the results of the recent experiment of Ref. \cite{kondo_exp}. In our previous work \cite{renkondo2}, we have analyzed this experiment by calculating the intensities of effective 1D spin-exchange interaction between the $e$- and $g$-atoms, and in the calculation, the axial lattice potential for the $g$-atom was ignored. In the present manuscript, we can include this optical lattice. More importantly, in comparison with the effective interaction intensity, the spin-exchange collision rate $R_{\rm se}$ (i.e., the ``cross-section" of the spin-exchange collision) is much more directly related to the experimental observations of Ref \cite{kondo_exp} on the amount of the atoms whose spin are flipped during the scattering.
Thus, in the current work, we calculate $R_{\rm se}$ for the experimental system. Since $R_{\rm se}$ depends on the incident quasi-momentum of the $g$-atom, by comparing the theoretical and experimental results we estimate this quasi-momentum and investigate the finite-quasi-momentum effect of this experiment. Our results show that this effect is very significant. Explicitly, 
the resonance position is very sensitive to the incident quasi-momentum of the $g$-atom. 

The remainder of this paper is organized as follows. We
describe the detail of our quasi-(1+0)D system in Sec. II and show the forms of the corresponding tight-binding models in Sec. III. In Sec. IV we calculate the exact low-energy scattering amplitudes between the $e$- and $g$-atoms and derive the interaction parameters for the tight-binding models. The applicability of the ``projection approximation" and the experimental cases for the $^{171}$Yb and $^{173}$Yb atoms are also analyzed in this section. In Sec. V we calculate the spin-exchange collision rate for the experimental systems of Ref. \cite{kondo_exp}. A summary of our results and some discussions
are given in Sec. VI. In the appendixes, we present details of our calculation.


\section{System and Hamiltonian}

As mentioned in Sec. I, we consider two alkaline-earths (like) atoms of the same species, which are in the electronic $^1$S$_0$ ($g$-) and $^3$P$_0$ ($e$-) state, respectively. Also, each atom can be in two nuclear-spin sates, which are denoted as $\uparrow$ and $\downarrow$ (Fig.~1(a)). We focus on the case with zero magnetic fields ($B=0$) where these two spin states are degenerate. Furthermore, as shown in Fig.~1(b), the two atoms are confined in a 2D isotropic harmonic potential in the $x$-$y$ plane (i.e., a quasi-1D tube along the $z$-direction), which has the same frequency for both of the two atoms. Also, in the $z$-the direction the $g$-atom experiences an optical lattice and the $e$-atom is localized by another 1D harmonic trapping potential.

The Hamiltonian for our two-body problem is given by
\begin{align}
H=H_0+U,\label{h}
\end{align}
where $H_0$ and $U$ are the free Hamiltonian and inter-atomic interaction, respectively. Explicitly, $H_0$ can be expressed as
\begin{align}
H_{0}=H_{\perp}+H_{z}^{(g)}+H_{z}^{(e)}.\label{H0}
\end{align}
Here
\begin{align}
H_{\perp}=-\frac{\hbar^{2}}{2\mu}\nabla_{\bm \rho}^{2}+\frac{1}{2}\mu\omega_{\perp}^{2}\rho^{2}\label{hperp}
\end{align}
is the Hamiltonian for the relative motion of the two atoms in the transverse directions ($x$-$y$ plane), with ${\bm \rho}$ being the relative coordinate vector on the transverse direction, $\mu$ being the two-atom reduced mass, and $\omega_\perp$ being the transverse confinement frequency.
We would like to emphasis that since the confinement in the $x$-$y$ plane is a harmonic potential with the same frequency for the two atoms, in this plane the 
center-of-mass motion of these two atoms can be separated from the relative motion, and is ignored in our Hamiltonian. On the other hand, in Eq.~(\ref{H0}) the terms $H_{z}^{(g)}$ and $H_{z}^{(e)}$ are the Hamiltonians for the motion of the $g$-atom and $e$-atom along the $z$-direction, respectively. They can be expressed as
\begin{eqnarray}
H_{z}^{(g)}&=&-\frac{\hbar^{2}}{2m}\frac{d^{2}}{dz_{g}^{2}}+s_{g}E_{R}\sin^{2}\left(k_{0}z_{g}\right)\label{hzg};\\
H_{z}^{(e)}&=&-\frac{\hbar^{2}}{2m}\frac{d^{2}}{dz_{e}^{2}}+\frac{1}{2}m\omega_{z}^{2}z_e^{2},\label{hze}
\end{eqnarray}
where $m=2\mu$ is the single-atom mass, $z_{g(e)}$ is the $z$-coordinate of the $g$- ($e$-) atom, $k_{0}$ is the wave vector of laser for the optical lattice of the $g$-atom, $s_g$ and $E_{R}=\hbar^{2}k_0^2/(2m)$ are the depth and the corresponding recoil energy of this lattice, respectively, and $\omega_z$ is the trapping frequency for the $e$-atom in the $z$-direction.

Moreover, as shown in previous researches (e.g., Ref. \cite{kondo_exp,Yb173as,Yb171,spin-exchange1,spin-exchange2,XiboSUN}), the inter-atomic interaction $U$ for our system is diagonal in the basis of nuclear-spin singlet and triplet states, and can be expressed as 
\begin{equation}
U=U_{+}{\cal P}_++U_{-}{\cal P}_-,\label{vv}
\end{equation}
where ${\cal P}_{+(-)}$ is the projection operator for the spin-singlet (triplet) sates, and can be written as
\begin{eqnarray}
{\cal P}_+=|+\rangle\langle+|,\ \ {\cal P}_-=\sum_{q=0,\pm1}|-,q\rangle\langle-,q|.\label{pj}
\end{eqnarray}
Here 
\begin{eqnarray}
|+\rangle & = & \frac{1}{\sqrt{2}}\left(|\uparrow\rangle_{g}|\downarrow\rangle_{e}-|\downarrow\rangle_{g}|\uparrow\rangle_{e}\right)\label{p}
\end{eqnarray}
is the spin-singlet state and 
\begin{eqnarray}
|-,0\rangle & = & \frac{1}{\sqrt{2}}\left(|\uparrow\rangle_{g}|\downarrow\rangle_{e}+|\downarrow\rangle_{g}|\uparrow\rangle_{e}\right);\label{m0}\\
|-,+1\rangle & = & |\uparrow\rangle_{g}|\uparrow\rangle_{e};\label{m1}\\
|-,-1\rangle & = & |\downarrow\rangle_{g}|\downarrow\rangle_{e},\label{mn1}
\end{eqnarray}
are the triplet states. In Eq. (\ref{vv}) $U_{+(-)}$ is the Huang-Yang pseudo potential corresponding to the singlet (triplet) states, and takes to the following form
\begin{equation}
U_{\pm}= \frac{2\pi\hbar^{2}a_s^{(\pm)}}{\mu}\delta({\bf r})\frac{\partial}{\partial r}\left(r\cdot\right),\label{hy}
\end{equation}
where $a_s^{(\pm)}$ are the corresponding scattering lengths, ${\bf r}={\bm \rho}+(z_g-z_e){\bf e}_z$ is the relative-position vector of the two atoms and $r=|{\bf r}|$, with ${\bf e}_z$ being the unit vector along the $z$-direction.
For the realistic alkali-earth (like) atoms, we usually have $a_s^{(+)}\neq a_s^{(-)}$, which is inaccordance with the existence of the spin-exchange interaction. For instance,
we have $a_s^{(+)}\approx1878a_{0}$ and $a_s^{(-)}\approx216a_{0}$
for $^{173}$Yb atoms \cite{Yb173as,Fallani-PRX} , and $a_s^{(+)}\approx 225a_{0}$ and $a_s^{(-)}\approx 355a_{0}$
for $^{171}$Yb atoms \cite{Yb171}.

\section{Tight-binding models}

\subsection{2-body tight-binding models}

Now we consider the case where 
the temperature $T$ of our system is low enough so that $k_BT$ is much smaller than $\hbar\omega_z$, $\hbar\omega_\perp$, as well as the minimum energy of the second band of the axial optical lattice for the $g$-atom, with $k_B$ being the Boltzmann constant. 
We further assume that the axial lattice is deep enough so that only the hopping between the nearest sites is required to be taken into account. 
In this case, when the two atoms are separated from each other, the transverse relative motion is frozen in the ground state of the transverse confinement, and the motion of the $g$-atom and $e$-atom in the $z$-direction are frozen in the lowest band of the optical lattice and the ground state of the axial harmonic trap, respectively. As a result, the system can be described by 1D tight-binding models, where only the hopping of the $g$-atom between the nearest sites of the axial lattice and the nuclear spin of these two atoms are taken into account. Thus, in these models the free Hamiltonian of the $g$-atom can be expressed as 
\begin{eqnarray}
T_{\rm 1D}=-t\sum^{\infty}_{n=-\infty}\left( |n\rangle_g\langle n+1|+{\rm h.c.}\right),\label{t1d}
\end{eqnarray}
where $|n\rangle_g$ is the Wannier state of the $g$-atom for the $n$-th site of the axial optical lattice, and $t$ is the hopping rate. Explicitly, the wave function of state $|n\rangle_g$ is
$
_g\langle z_g|n\rangle_g=w(z_g-nl_0)
$
with $|z_g\rangle_g$ being the eigen-state of the $z$-coordinate of the $g$-atom and $w(z)$ being the Wannier function of the axial optical lattice, and
\begin{eqnarray}
l_0=\frac{\pi}{k_0}\label{l0}
\end{eqnarray}
is the lattice constant. The hopping rate $t$ can be expressed as
\begin{align}
t=&\int dzw(z)\left[\frac{\hbar^{2}}{2m}\frac{d^{2}}{dz^{2}}-s_{g}E_{R}\sin^{2}(k_{0}z)\right]w(z-l_0).
\end{align}

Now we consider the interaction term in the tight-binding models. We notice that the following three facts are important for the analysis of this problem:

\begin{itemize}
\item[{\bf a)}] The effective interaction should be able to reproduce the low-energy scattering amplitude between the $g$- and $e$-atom. \\

\item[{\bf b)}] Our system is invariable under the total spatial reflection $z_{g}\to -z_{g}, z_{e}\to -z_{e}$ of the two atoms. As a result, the total spatial parity 
 $\mathbb{P}$
of the two atoms is conserved. Therefore, there are two partial-wave scattering channels, i.e., the even-wave and odd-wave which correspond to $\mathbb{P}=+1$ and $\mathbb{P}=-1$, respectively. \\
 
\item[{\bf c)}] For many realistic quasi-(1+0)D systems, 
the odd-wave scattering is not negligible only when the system is under an odd-wave scattering resonance. Otherwise, the odd-wave scattering amplitude can be safely ignored.
\end{itemize}
Considering the above facts, we find that we may have the following two tight-binding models for our system, with 
different effective interactions corresponding to two different cases:

\subsubsection*{\rm \bf Model (I): For the cases where the odd-wave scattering can be ignored.}

 As shown in Fig.~1(c), in this model the $e$-atom, which is confined at the 0-th site of the axial lattice (i.e., the region around $z_e=0$), only interacts with the $g$-atom in the same site. Accordingly,
the effective interaction is described as a local potential:
\begin{eqnarray}
U_{\rm 1D}^{\rm (I)}&\equiv&\sum_{\xi=+,-}u_{0,\xi}^{\rm (I)}|0\rangle_g\langle 0|{\cal P}_\xi,\label{u1di}
\end{eqnarray}
where the projection operators ${\cal P}_\xi$ ($\xi=+,-$) are defined in Eq. (\ref{pj}). 
Thus, the complete tight-binding Hamiltonian for this model is
\begin{eqnarray}
H_{\rm 1D}^{\rm (I)}\equiv T_{\rm 1D}+U_{\rm 1D}^{\rm (I)}.\label{h1di}
\end{eqnarray}

Straightforward calculations show that the odd-wave scattering amplitude corresponding to $H_{\rm 1D}^{\rm (I)}$ is exactly zero. Therefore, this model can be used for the cases where the odd-wave scattering amplitude is negligible. On the other hand, this model should be able to re-produce the correct even-wave interaction scattering amplitudes. Therefore, matching the eve-wave scattering amplitudes given by $H_{\rm 1D}^{\rm (I)}$ with the ones corresponding to the exact Hamiltonian $H$ of Eq. (\ref{h}), we can derive the values of $u_{0,\pm}^{\rm (I)}$. The detailed calculations are given in the next section.


\subsubsection*{\rm\bf Model (II): For the cases where the odd-wave scattering cannot be neglected.}

As shown in Fig.~1(d), in this model the $e$-atom interacts with not only the $g$-atom in the same site (the 0-th site), but also the $g$-atom in the nearest neighbor sites (the $\pm 1$-st sites). Thus, the 
effective interaction potential can be expressed as
\begin{align}
U_{\rm 1D}^{\rm (II)}&\equiv\sum_{\xi=\pm}\left[u_{0,\xi}^{\rm (II)}|0\rangle_g\langle 0|+u_{1,\xi}^{\rm (II)}\sum_{i=\pm1}|i\rangle_g\langle i|\right]{\cal P}_\xi,
\nonumber\\
\label{u1dii}
\end{align}
and the total tight-binding Hamiltonian for this case is 
\begin{eqnarray}
H_{\rm 1D}^{\rm (II)}\equiv T_{\rm 1D}+U_{\rm 1D}^{\rm (II)}.\label{h1dii}
\end{eqnarray}

 Due to the second term on the right-hand side of Eq. (\ref{u1dii}), both the even- and odd-wave scattering amplitudes corresponding to $H_{\rm 1D}^{\rm (II)}$ are non-zero. Therefore, this model
 can re-produce the explicit scattering amplitudes in both of the two partial wave channels, and thus
 can be used for the general cases, especially the cases where the odd-wave scattering of our system cannot be ignored.

 Similar as in Model (I), one can determine the interaction parameters $u_{0,\pm}^{\rm (II)}$ and $u_{1,\pm}^{\rm (II)}$ by matching the scattering amplitudes 
for $H_{\rm 1D}^{\rm (II)}$ 
with the ones for the exact Hamiltonian $H$ of Eq. (\ref{h}). The detailed calculations are also given in the next section. 
 

At the end of this subsection, we emphasize that in principle the model (II) can also be used for the cases that the odd-wave scattering amplitude is negligible, as the model (I). In these cases, both the even- and odd-wave low-energy scattering amplitudes of model (II) are the same as the ones for the explicit Hamiltonian $H$, although the odd-wave ones are very small. Nevertheless, as shown above, the model (II) is more complicated than
model (I) in which the odd-wave scattering is ignored.
 
\subsection{Many-body tight-binding models}

We can directly generalize the above two tight-binding models to the systems where there are many $g$-atoms moving in the axial lattice, interacting with a single $e$-atom fixed at the 0-th site. To this end, we re-write the Hamiltonians $H_{\rm 1D}^{\rm (I)}$ and $H_{\rm 1D}^{\rm (II)}$ for these two models in the second-quantized form as
\begin{align}
H_{\rm MB}^{\rm (I)}=&-t\sum^{\infty}_{n=-\infty}\sum_{\sigma=\uparrow,\downarrow}\left(c^{\dagger}_{n+1,\sigma}c_{n,\sigma}+{\rm h.c.}\right)\nonumber\\
&+\Omega_{0}^{\rm (I)}\left(\frac{1}{2}\sigma_{z}^{(g,0)}\sigma_{z}^{(e)}+\sigma_{+}^{(g,0)}\sigma_{-}^{(e)}+\sigma_{-}^{(g,0)}\sigma_{+}^{(e)}\right)\nonumber\\
&+\Lambda_{0}^{\rm (I)} n^{(g,0)}\label{hmb1}
\end{align}
and
\begin{align}
H_{\rm MB}^{\rm (II)}=&-t\sum^{\infty}_{n=-\infty}\sum_{\sigma=\uparrow,\downarrow}\left(c^{\dagger}_{n+1,\sigma}c_{n,\sigma}+{\rm h.c.}\right)\nonumber\\
&+\Omega_{0}^{\rm (II)}\left(\frac{1}{2}\sigma_{z}^{(g,0)}\sigma_{z}^{(e)}+\sigma_{+}^{(g,0)}\sigma_{-}^{(e)}+\sigma_{-}^{(g,0)}\sigma_{+}^{(e)}\right)\nonumber\\
&+\Omega_{1}^{\rm (II)}\sum_{j=\pm1}\left(\frac{1}{2}\sigma_{z}^{(g,j)}\sigma_{z}^{(e)}+\sigma_{+}^{(g,j)}\sigma_{-}^{(e)}+\sigma_{-}^{(g,j)}\sigma_{+}^{(e)}\right)\nonumber\\
&+\Lambda_{0}^{\rm (II)} n^{(g,0)}+\Lambda_{1}^{\rm (II)}\sum_{j=\pm1} n^{(g,j)},\label{hmb2}
\end{align}
respectively. Here $c_{n,\sigma}$ and $c^\dagger_{n,\sigma}$ ($\sigma=\uparrow,\downarrow$) are the annihilation and creation operators of $g$-atom at the $n$-th site of the axial lattice with nuclear spin $\sigma$, respectively, and the operators $\sigma_{z,\pm}^{(g,j)}$ and $n^{(g,j)}$ are defined as
\begin{eqnarray}
\sigma_{z}^{(g,j)}&=&c^{\dagger}_{j,\uparrow}c_{j,\uparrow}-c^{\dagger}_{j,\downarrow}c_{j,\downarrow},\\
\sigma_{+}^{(g,j)}&=&c^{\dagger}_{j,\uparrow}c_{j,\downarrow},\\
\sigma_{-}^{(g,j)}&=&c^{\dagger}_{j,\downarrow}c_{j,\uparrow},\\
n^{(g,j)}&=&c^{\dagger}_{j,\uparrow}c_{j,\uparrow}+c^{\dagger}_{j,\downarrow}c_{j,\downarrow},
\end{eqnarray}
In addition, $\sigma_{z}^{(e)}=|\uparrow\rangle_{e}\langle\uparrow|-|\downarrow\rangle_{e}\langle\downarrow|$, $\sigma_{+}^{(e)}=|\uparrow\rangle_{e}\langle\downarrow|$ and $\sigma_{-}^{(e)}=|\downarrow\rangle_{e}\langle\uparrow|$ are the spin operators of the fixed $e$-atom. The interaction strengths $\Omega_{\alpha}^{(\beta)}$ and $\Lambda_{\alpha}^{(\beta)}$ ($\alpha=0$ for $\beta={\rm I}$ and $\alpha=0, 1$ for $\beta={\rm II}$) are defined as follows,
\begin{align}
\Omega_{\alpha}^{(\beta)}=&\frac{u_{\alpha,-}^{(\beta)}-u_{\alpha,+}^{(\beta)}}{2}\label{Omega};\\
\Lambda_{\alpha}^{(\beta)}=&\frac{3u_{\alpha,-}^{(\beta)}+u_{\alpha,+}^{(\beta)}}{4}.\label{Lambda}
\end{align}
The parameters $t$, $u_{0,\pm}^{\rm (I)}$ and $u_{0,\pm}^{\rm (II)}$ and $u_{1,\pm}^{\rm (II)}$ are exactly the ones introduced in the last subsection. 

It is easy to prove that when there is only one $g$-atom, the Hamiltonians $H_{\rm MB}^{\rm (I)}$ and $H_{\rm MB}^{\rm (II)}$ are mathematically equivalent with $H_{\rm 1D}^{\rm (I)}$ and $H_{\rm 1D}^{\rm (II)}$, respectively. In addition, $H_{\rm MB}^{\rm (I)}$ and $H_{\rm MB}^{\rm (II)}$ can also be used for the cases with more than one $g$-atoms. Here we also emphasis that, since the interaction parameters in the models (I) and (II) are determined by the scattering amplitude between a single $g$-atom and the localized $e$-atom, these models should be used in the cases where the $g$-atoms are dilute enough so that the probability of the three-body collision between two $g$-atoms in different spin states and the $e$-atom can be ignored, e.g., the cases where the 1D density of the $g$-atoms is much less than $1/l_0$, with the lattice constant $l_0$ being defined in Eq. (\ref{l0})

\label{sec:sa}
\section{Scattering Amplitude and the $u$-parameters}

As shown in the above section, the interaction parameters in the above two tight-binding models are determined by the low-energy scattering amplitudes of the explicit Hamiltonian $H$. In addition, since both $H$ and the tight-binding Hamiltonians $H_{\rm 1D}^{\rm (I,II)}$ of our system are diagonal in the basis of spin-singlet and triplet states, these two spin channels are independent of each other. As a result, the interaction parameters $\{u_{0,+}^{\rm (I)}, u_{0,+}^{\rm (II)}, u_{1,+}^{\rm (II)}\}$ and $\{u_{0,-}^{\rm (I)}, u_{0,-}^{\rm (II)}, u_{1,-}^{\rm (II)}\}$ are determined by the scattering amplitudes of the Hamiltonians $H_0+U_+$ and $H_0+U_-$, respectively, with $H_0$ and $U_\pm$ being defined in Eqs. (\ref{H0}, \ref{hy}).

In this section, we calculate these amplitudes and the interaction parameters. We will show that these $u$-parameters can be efficiently controlled by the confinement and lattice potentials via the CIRs of our system.

\subsection{Explicit low-energy scattering amplitudes}

Now we calculate the low-energy scattering amplitudes for the Hamiltonian 
\begin{eqnarray}
H_\xi\equiv H_0+U_\xi \ \ (\xi=+,-).\label{hxi}
\end{eqnarray}
The incident state $ \Psi_{\rm in}({\bm \rho},z_{g},z_{e})$ is an eigen-state of the free Hamiltonian $H_0$, with corresponding eigen-energy (i.e., the scattering energy) $E$.
$ \Psi_{\rm in}({\bm \rho},z_{g},z_{e})$ can be expressed as
\begin{eqnarray}
 \Psi_{\rm in}({\bm \rho},z_{g},z_{e})=\chi_0({\bm \rho})\phi_{0}(z_{e})\psi(z_{g}),\label{psiin}
\end{eqnarray}
where $\chi_0({\bm \rho})$ and $\phi_{0}(z_{e})$ are the ground states of the free Hamiltonians $H_\perp$ and $H_z^{(e)}$ for the transverse relative motion and the $e$-atom axial motion, respectively. In Eq. (\ref{psiin}), $\psi(z_{g})$ is the lowest band Bloch wave function of the $g$-atom. It is an eigen-state of the free Hamiltonian $H_z^{(g)}$ for the axial motion of the $g$-atom, and satisfies
\begin{align}
\label{hzge}
H_z^{(g)}\psi(z_{g})={\cal E}\psi(z_{g}),
\end{align}
with ${\cal E}=E-\hbar\omega_{\perp}-\hbar\omega_{z}/2$. Using the expression (\ref{hzg}) of $H_z^{(g)}$, we find that Eq. (\ref{hzge}) can be mathematically reduced to the Mathieu equation \cite{Mathiue},
\begin{align}
\label{me}
\left\{\frac{d^{2}}{dz_{g}^{2}}+\left[a-2q\cos(2k_{0}z_{g})\right]\right\}\psi(z_g)=0,
\end{align}
with indexes $a=2m{\cal E}/\hbar^{2}k_{0}^{2}-s_{g}/2$ and $q=-s_g/4$.
Thus, $\psi(z_{g})$ can be expressed as
\begin{align}
\label{bloch}
\psi(z_{g})=e^{ikz_{g}}{\cal F}(z_{g}).
\end{align}
Here the function ${\cal F}(z_g)$ can be further expressed as ${\cal F}(z_g)={\cal M}(k_0z_g)$, with ${\cal M}(s)$ being the Mathieu function of the first kind, which satisfies ${\cal M}(s)= {\cal M}(s+\pi)$. In Eq. (\ref{bloch}), the parameter $k$ is the characteristic exponent of the Mathieu equation (\ref{me}). It is a function of ${\cal E}$ and satisfies $k>0$. 
Apparently, ${\cal F}(s)$ and $k$ are also the Bloch wave function and
 the positive quasi-momentum corresponding to the eigen-energy ${\cal E}$ of the lattice Hamiltonian $H_z^{(g)}$, respectively. Using the properties of the Mathieu function, it can be proved that in the absence of the axial lattice (i.e., $s_g=0$), Eq. (\ref{bloch}) becomes $\psi_{k}(z_{g})=e^{ikz_g}$, i.e., $\psi_{k}(z_{g})$ returns to the plane-wave state corresponding to momentum $k$.

Furthermore, the scattering state $\Psi_{\xi}(\bm{\rho},z_{e},z_{g})$ ($\xi=+,-$) of our problem
is determined by the Schr$\ddot{{\rm o}}$dinger equation 
\begin{align}
\label{se}
(H_{0}+U_{\xi})\Psi_{\xi}(\bm{\rho},z_{e},z_{g})=E\Psi_{\xi}(\bm{\rho},z_{e},z_{g}), 
\end{align}
as well as the out-going boundary condition in the limit $|z_{g}|\rightarrow\infty$, which can be expressed as
\begin{widetext}
\begin{eqnarray}
 \Psi_{\xi}(\bm{\rho},z_{e},|z_{g}|\rightarrow\infty)= \chi_0({\bm \rho})\phi_{0}(z_{e})\left[
 e^{ikz_{g}}{\cal F}(z_{g})+f^{\rm (e)}_\xi(k) e^{ik|z_{g}|}{\cal F}(|z_{g}|)+
 f^{\rm (o)}_\xi(k){\rm sign}(z_g)e^{ik|z_{g}|}{\cal F}(|z_{g}|)
 \right]. \label{psi2}
\end{eqnarray}
Here $f^{\rm (e)}_\xi(k)$ and $f^{\rm (o)}_\xi(k)$ are the even-wave and odd-wave scattering amplitudes, respectively. It is clear that Eq. (\ref{psi2}) can be re-written in an intuitive form 
\begin{eqnarray}
 \Psi_{\xi}(\bm{\rho},z_{e},|z_{g}|\rightarrow\infty)= \chi_0({\bm \rho})\phi_{0}(z_{e})\times\left\{
 \begin{array}{lll}
 e^{ikz_{g}}{\cal F}(z_{g})+r_\xi (k)e^{-ikz_{g}}{\cal F}(-z_{g})&&({\rm for}\ z_g\rightarrow-\infty)\\
 t_\xi (k)e^{ikz_{g}}{\cal F}(z_{g})&&({\rm for}\ z_g\rightarrow+\infty)
 \end{array}
 \right.,\label{newpsi}
\end{eqnarray}
\end{widetext}
where $r_\xi(k)$ and $t_\xi(k)$ are the reflection and transmission amplitudes, respectively, and are related to $f_{\xi}^{{\rm (e/o)}}(k)$ via
$
r_\xi(k)=f_\xi^{\rm (e)}(k)-f_\xi^{\rm (o)}(k)
$ and $
t_\xi(k)=f_\xi^{\rm (e)}(k)+f_\xi^{\rm (o)}(k)+1$.

Furthermore, in the low-energy limit, i.e., $k\to0$, the scattering amplitudes can be expressed as
\begin{align}
f_\xi^{\rm (e)}(k)&=-\frac{1}{1+ika_\xi^{\rm (e)}+{\cal O}(k^{3})};\label{ae}\\
f_\xi^{\rm (o)}(k)&=-\frac{ik}{ik+\frac{1}{a_\xi^{\rm (o)}}+{\cal O}(k^{2})}\label{ao},
\end{align}
where $a_\xi^{\rm (e)}$ and $a_\xi^{\rm (o)}$ are defined as the even- and odd-wave 1D scattering length, respectively. These two scattering lengths describe the significance of the low-energy scattering effect. Explicitly, the even- and odd-wave scattering is significant in the limits $a_\xi^{\rm (e)}\rightarrow 0$ and $a_\xi^{\rm (o)}\rightarrow \infty$, respectively.

We solve the Schr$\ddot{{\rm o}}$dinger equation (\ref{se}) with the out-going boundary condition and calculate the scattering amplitudes $f_{\xi}^{{\rm (e/o)}}(k)$ and the scattering lengths $a_{\xi}^{{\rm (e/o)}}$. The details of our calculation are shown in Appendix \ref{sec:green}. Our method is a generalization of our previous work \cite{renkondo2} for the systems without the axial optical lattice. The key step of this generalization is the calculation of the Green's function in the aid of the Mathieu function.

For our system, $a_{\xi}^{{\rm (e/o)}}$ ($\xi=+,-$) depends not only on the 3D scattering length $a_s^{(\xi)}$, but also on the parameters of the axial lattice and the 
 confinement potentials, i.e., $s_g$, $l_0$, and
\begin{eqnarray}
a_{\perp} \equiv\sqrt{\frac{2\hbar}{m\omega_\perp}},\ a_{z} \equiv\sqrt{\frac{\hbar}{m\omega_z}}.\label{ccll}
\end{eqnarray}
When these parameters are tuned to some particular values, we may have $a_\xi^{\rm (e)}=0$ or $a_\xi^{\rm (o)}= \infty$. Namely, the even- or odd-wave scattering can be resonantly enhanced. These effects are the even-wave or odd-wave CIR, respectively.
 As shown in previous works \cite{Olshanii1998,Bergeman2003}, these CIRs essentially result from the virtual transitions to the excited 
states of $H_\perp$, $H_z^{(e)}$ and the excited bands of $H_z^{(g)}$, which occur during the scattering process and are explicitly taken into account in our calculation.

\begin{figure}
\centering
\includegraphics[width=0.3\textwidth]{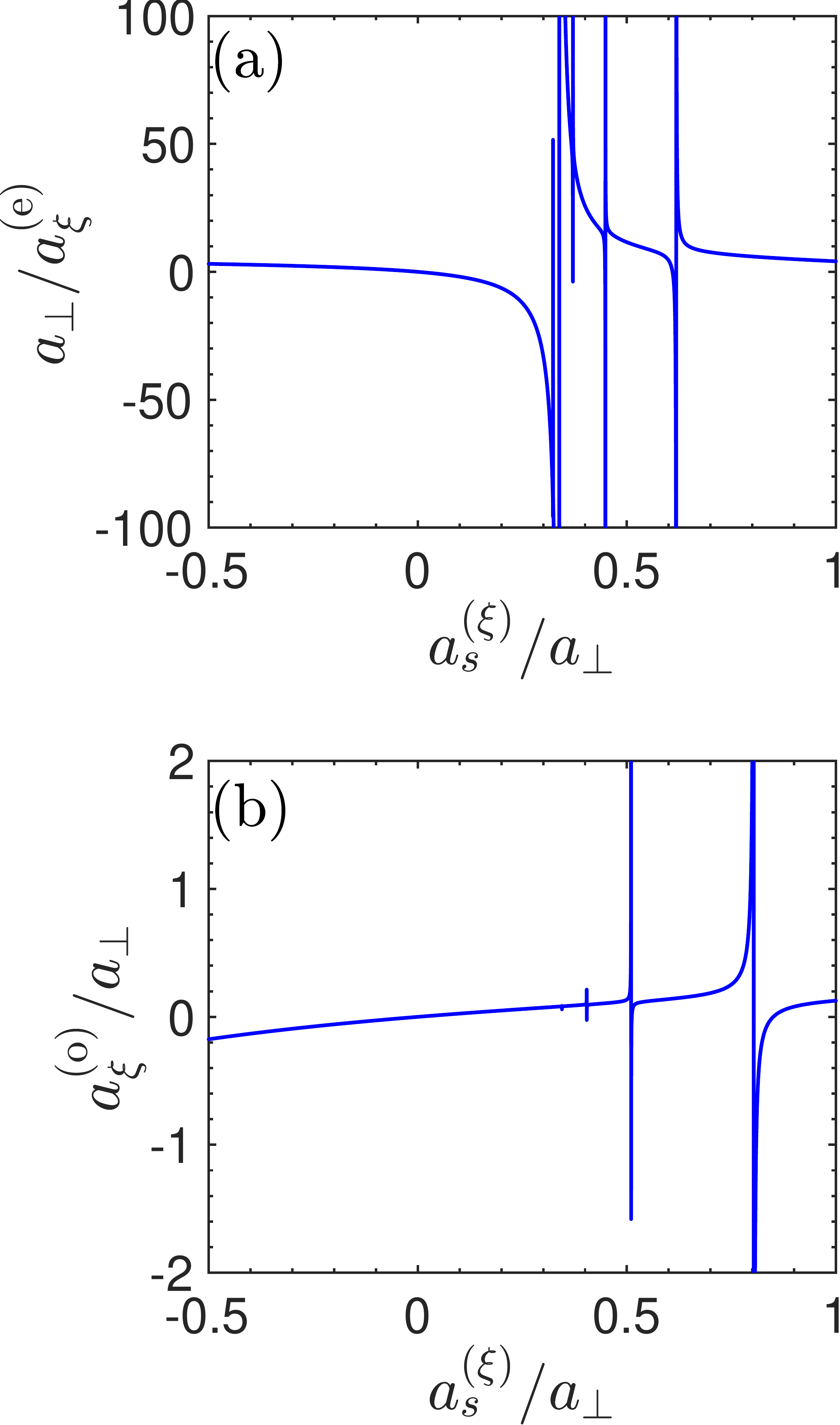} 
\caption{(color online) The reciprocal of the even-wave scattering length $a^{\rm (e)}_\xi$ {\bf (a)} and the odd-wave scattering length $a^{\rm (o)}_\xi$ {\bf (b)} as functions of $a_s^{(\xi)}/a_\perp$ for the cases with $a_{z}/a_{\perp}=0.8$, $a_z/l_0=0.15$, and $s_{g}=5$. As shown in the main text, $1/a^{\rm (e)}_\xi\to\infty\ (a^{\rm (o)}_\xi\to\infty)$ indicates the presence of a even-wave (odd-wave) CIR.\label{scatteringamplitude}}
\end{figure}

In Fig.~\ref{scatteringamplitude}(a) and (b), we show $1/a_{\xi}^{{\rm (e)}}$ and $a_{\xi}^{{\rm (o)}}$ for the typical cases with $a_z/l_0=$0.15, $a_{z}/a_{\perp}=0.8$ and $s_{g}=5$. It is clearly shown that multi-CIRs can appear for both even-wave and odd-wave scattering. That is essentially due to the coupling between the center-of-mass and relative motion of these two atoms in the $z$-direction. 
Similar phenomenon also occurs for other mixed-dimensional systems \cite{yvan,Nishida,shina,mixed_exp}.
In addition, 
Fig.~\ref{scatteringamplitude} shows that the even-wave CIRs are much broader than the odd-wave CIRs (notice that the range of the vertical axis of Fig.~\ref{scatteringamplitude}(a) is $[-100,+100]$, while the one of Fig.~\ref{scatteringamplitude}(b) is $[-2,+2]$). That is consistent with the fact that in the low-energy limit the odd-wave scattering effect is much weaker than the even-wave scattering (i.e., $\lim_{k\rightarrow 0}f_\xi^{\rm (o)}(k)=0$ while $\lim_{k\rightarrow 0}f_\xi^{\rm (e)}(k)=-1$).

\subsection{$u$-parameters in the tight-binding models}

\begin{figure*}[t]
\centering
\includegraphics[width=0.9\textwidth]{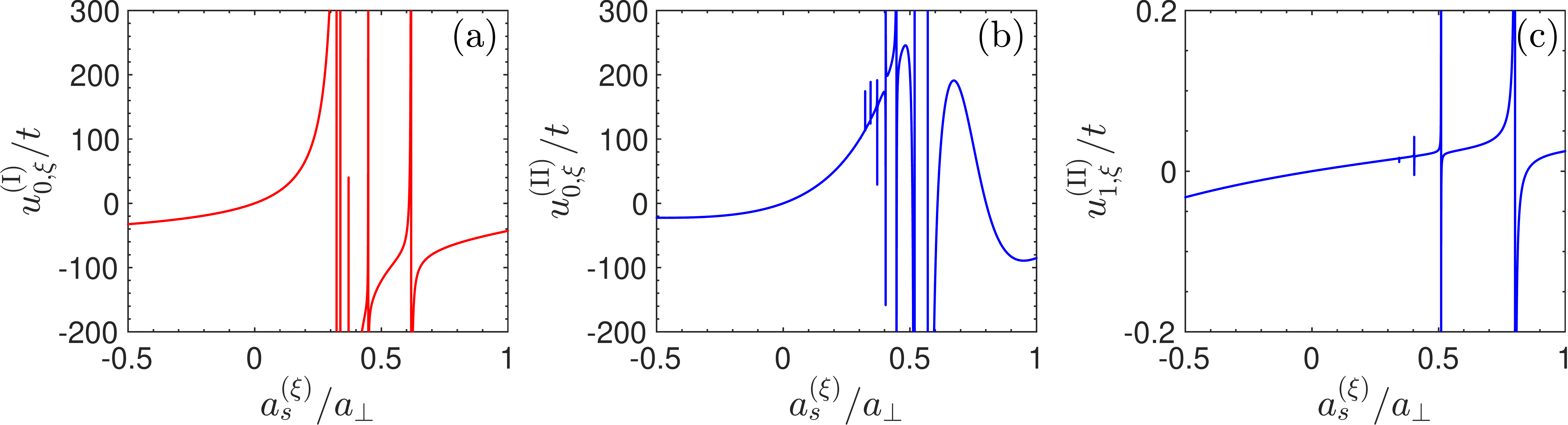} 
\caption{(color online) The $u$-parameters of the tight-binding models (I) and (II), as functions of the $s$-wave scattering length $a_s^{(\xi)}$. As in Fig.~(2), here we consider the cases with $a_{z}/a_{\perp}=0.8$, $a_z/l_0=0.15$ and $s_{g}=5$. \label{hubbardU}}
\end{figure*}

Now we consider the interaction parameters in the tight-binding models introduced in Sec. III, i.e, the parameters $u_{0,\xi}^{\rm (I)}, u_{0,\xi}^{\rm (II)}, u_{1,\xi}^{\rm (II)}$ ($\xi=+,-$). As shown above, these parameters should be determined by equalizing the low-energy scattering amplitudes given by the tight-binding Hamiltonians $H_{\rm 1D}^{\rm (I, II)}$ and the ones given by the exact Hamiltonian $H$, which are calculated in Sec. IV. A. Furthermore, as proved in Appendix \ref{sec:tightbinding}, in the low-energy limit the even- and odd-wave scattering amplitudes of $H_{\rm 1D}^{\rm (I, II)}$ in each spin channel also have the behaviors of Eqs. (\ref{ae}, \ref{ao}). Therefore, we can determine the $u$-parameters by equalizing the scattering lengths of the tight-binding models and the ones for the exact Hamiltonian $H$ (i.e., $a_\xi^{\rm (e/o)}$ obtained in the above subsection). According to straightforward calculations in Appendix \ref{sec:tightbinding}, this approach yields
that the interaction parameter $u_{0,\xi}^{\rm (I)}$ for the tight-binding model (I) of Sec. III. A is given by
\begin{eqnarray}
\frac{u_{0,\xi}^{\rm (I)}}{t}=-\frac{2l_0}{a_\xi^{\rm (e)}},\label{u0iii}
\end{eqnarray}
while the parameters $u_{0,\xi}^{\rm (II)}$ and $u_{1,\xi}^{\rm (II)}$ for the tight-binding model (II) are given by
\begin{align}
\frac{u_{0,\xi}^{\rm (II)}}{t}&=-\frac{2\left(l_0^{2}-2l_0a_\xi^{\rm (o)}+a_\xi^{\rm (e)}a_\xi^{\rm (o)}\right)}{l_0\left(a_\xi^{\rm (e)}-a_\xi^{\rm (o)}\right)}\label{hubbardU0};\\
\frac{u_{1,\xi}^{\rm (II)}}{t}&=\frac{a_\xi^{\rm (o)}}{l_0-a_\xi^{\rm (o)}}\label{hubbardU1},
\end{align}
respectively. Eqs.~(\ref{hubbardU0}) and (\ref{hubbardU1}) clearly show that 
when the odd wave interaction disappears, i.e., $a_{\xi}^{\rm (o)}=0$, 
we have $u_{1,\xi}^{\rm (II)}=0$ and $u_{0,\xi}^{\rm (II)}=u_{0,\xi}^{\rm (I)}$. Thus, in the absences of the odd-wave interaction our
model (II) exactly reduces to model (I).

Since $u_{0,\xi}^{\rm (I,II)}$ and $u_{1,\xi}^{\rm (II)}$ ($\xi=+,-$) are functions of the 1D scattering lengths $a_\xi^{\rm (e/o)}$, one can control these parameters by tuning the 3D scattering length $a_s^{(\xi)}$ and the confinement potentials via the CIRs, as shown in Fig.~\ref{hubbardU} where we illustrate $u_{0,\xi}^{\rm (I)}$, $u_{0,\xi}^{\rm (II)}$ and $u_{1,\xi}^{\rm (II)}$ for the same cases of Fig.~\ref{scatteringamplitude}.


\begin{figure}[t]
\centering
\includegraphics[width=0.4\textwidth]{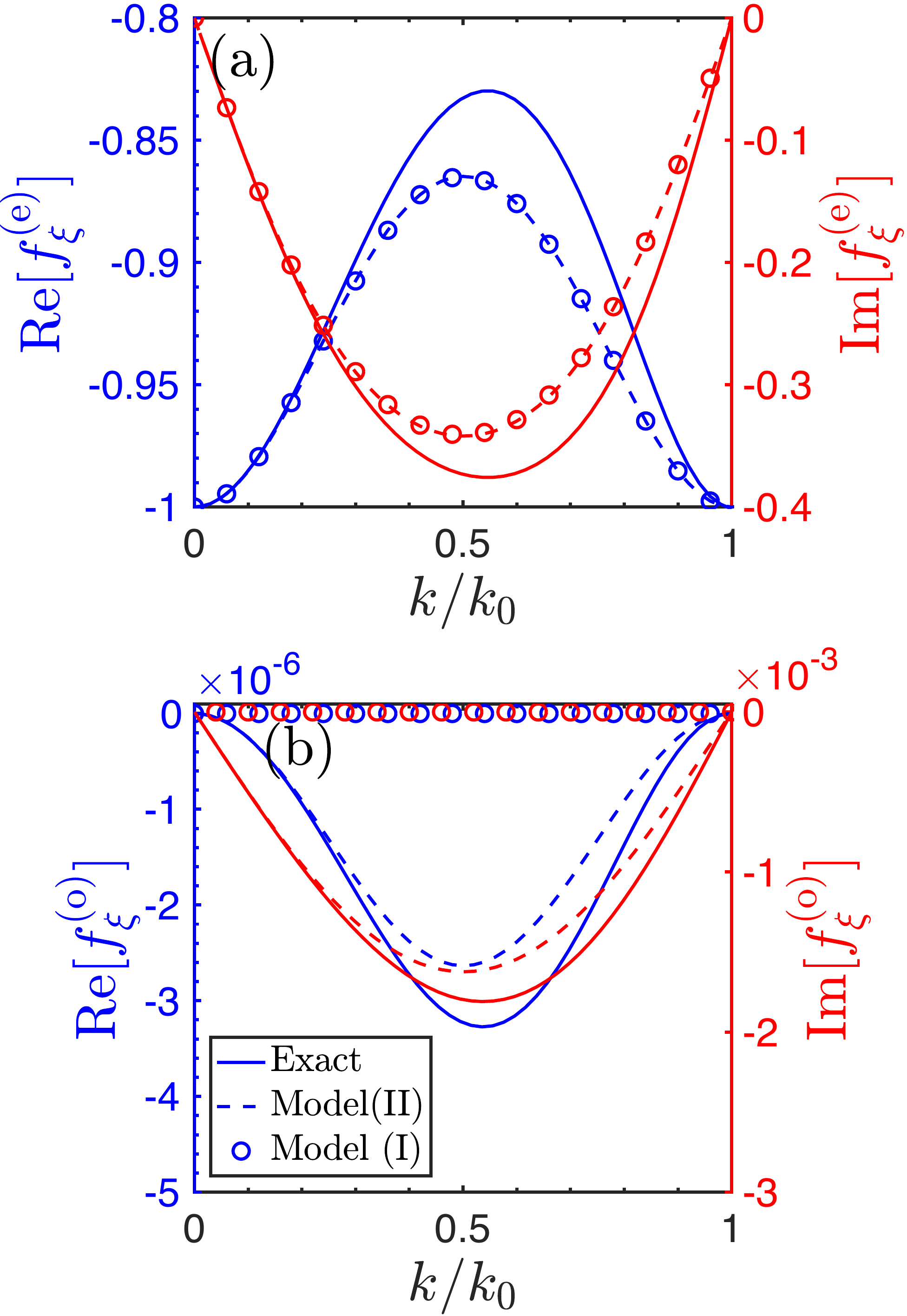} 
\caption{(color online) The real part (blue) and imaginary part (red) part of the even wave {\bf (a)} and odd wave {\bf (b)} scattering amplitudes, as functions of the quasi-momentum of the incident $g$-atom. 
We show 
the results given by the 
the exact Hamiltonian (the solid lines), the tight-binding model (I) (open circles) 
and the tight-binding model (II) (the dashed lines) for the case with $a_{z}/a_{\perp}=0.8$, $a_z/l_0=0.15$, $s_{g}=5$ and $a_s^{(\xi)}/a_\perp=0.032$.
 \label{fk}}
\end{figure}

In Fig.~\ref{fk} we further compare the scattering amplitudes of the tight-binding models with the above $u$-parameters and the ones of the exact Hamiltonian $H$ in Eq. (\ref{h}), for the cases with finite incident quasi momentum $k$ of the $g$-atom. We consider the system with $a_{z}/a_{\perp}=0.8$, $a_z/l_0=0.15$, $s_{g}=5$ and $a_s^{(\xi)}/a_\perp=0.032$ (corresponding to $u_{\xi,0}^{\rm (I)}/t=5.078$, $u_{\xi,0}^{\rm (II)}/t=5.037$, $u_{\xi,1}^{\rm (I)}/t=0.002$).
 In our problem the first Brillouin zone of the $g$-atom is $k\in[-\pi/l_0,+\pi/l_0]=[-k_0,+k_0]$. In Fig.~\ref{fk} we illustrate the results for $k\in[0,+k_0]$, because the scattering amplitudes for negative $k$ can be obtained via the relations $f_{\xi}^{(\rm e)}(k)=f_{\xi}^{(\rm e)}(-k)$ and $f_{\xi}^{(\rm o)}(k)=-f_{\xi}^{(\rm o)}(-k)$. 
 It is shown that the even-wave scattering amplitudes given by the two tight-binding models (I) and (II) and the one given by the exact Hamiltonian $H$ are almost the same for $k\lesssim 0.2$, and only have small quantitative differences (at most 0.05) for $k\gtrsim 0.2$. On the other hand, in this system, the explicit odd-wave scattering amplitude of $H$ is negligibly small (of the order of $10^{-4}$), and is almost the same as the result given by model (II) for $k\lesssim 0.2$. Therefore, the tight-binding models (I) and (II) can accurately reproduce the scattering amplitudes given by $H$ in a remarkable fraction of the first Brillouin zone.
 
Here we also emphasis that, since $a_s^{(+)}\neq
a_s^{(-)}$, the CIRs for channels with nuclear-spin singlet state (i.e., $\xi=+$) and the one with nuclear-spin triplet state (i.e., $\xi=-$) does not appear simultaneously.
Due to this fact, one can use the CIRs to resonantly enhance the spin-exchange parameters $\Omega_0^{\rm (I)}$ and $\Omega_{0,1}^{\rm (II)}$ in our tight-binding models (I) and (II). For instance, 
when an even-wave CIR with $\xi=+$ appears (i.e., when $a_+^{\rm (e)}=0$), we have $u_{0,+}^{\rm (I)}=\infty$ while $u_{0,-}^{\rm (I)}$ is still finite. According to Eq.~(\ref{Omega}), in this case $\Omega_0^{\rm (I)}$ can be resonantly enhanced. In addition, in the region of an even-wave CIR with $\xi=+$, there would be parameter point where $a_+^{\rm (e)}=a_+^{\rm (o)}$. At this point we have $u_{0,+}^{\rm (II)}=\infty$ while $u_{0,-}^{\rm (II)}$ is still finite, and thus $\Omega_0^{\rm (II)}$ can be resonantly enhanced. Similarly, when an odd-wave CIR with $\xi=+$ appears (i.e., $a_+^{\rm (o)}=\infty$), Eqs. (\ref{hubbardU1}) and (\ref{Omega}) yield that  $\Omega_1^{\rm (II)}$ of model (II) can be resonantly enhanced. It is clear that the above mechanism also works for the CIRs with $\xi=-$.

\subsection{Applicability of the projection approximation}

Now we consider the weakly-interacting cases where the 3D bare scattering length $a_s^{(\xi)}$ is small. For simplicity, here we focus on the case where the odd-wave scattering can be ignored and the system can be described by the tight-binding model (I).

In the limit $a_s^{(\xi)}\rightarrow 0$, during the scattering process the virtual transitions to the excited states and bands of the confinement and axial lattice potential can be neglected. As a result, we can make the ``projection approximation" where the interaction parameter $u_{0,\xi}^{\rm (I)}$ ($\xi=+,-$) of the tight-binding model (I) is approximated as the direct projection of the Huang-Yang pseudo potential $U_\xi$ on the ground states of the transverse confinement for the two-atom relative motion and the axial trapping potential of the $e$-atom, as well as the Wannier function of the 0-th site of the axial lattice of the $g$-atom, i.e.
\begin{align}
u_{0,\xi}^{\rm (I)}\approx u_{\rm project}\equiv2\hbar\omega_{\perp} a_{s}^{\rm (\xi)}\int dz|\phi_{0}(z)|^{2}|w(z)|^{2}.\label{app}
\end{align}

\begin{figure}[t]
\centering
\includegraphics[width=0.3\textwidth]{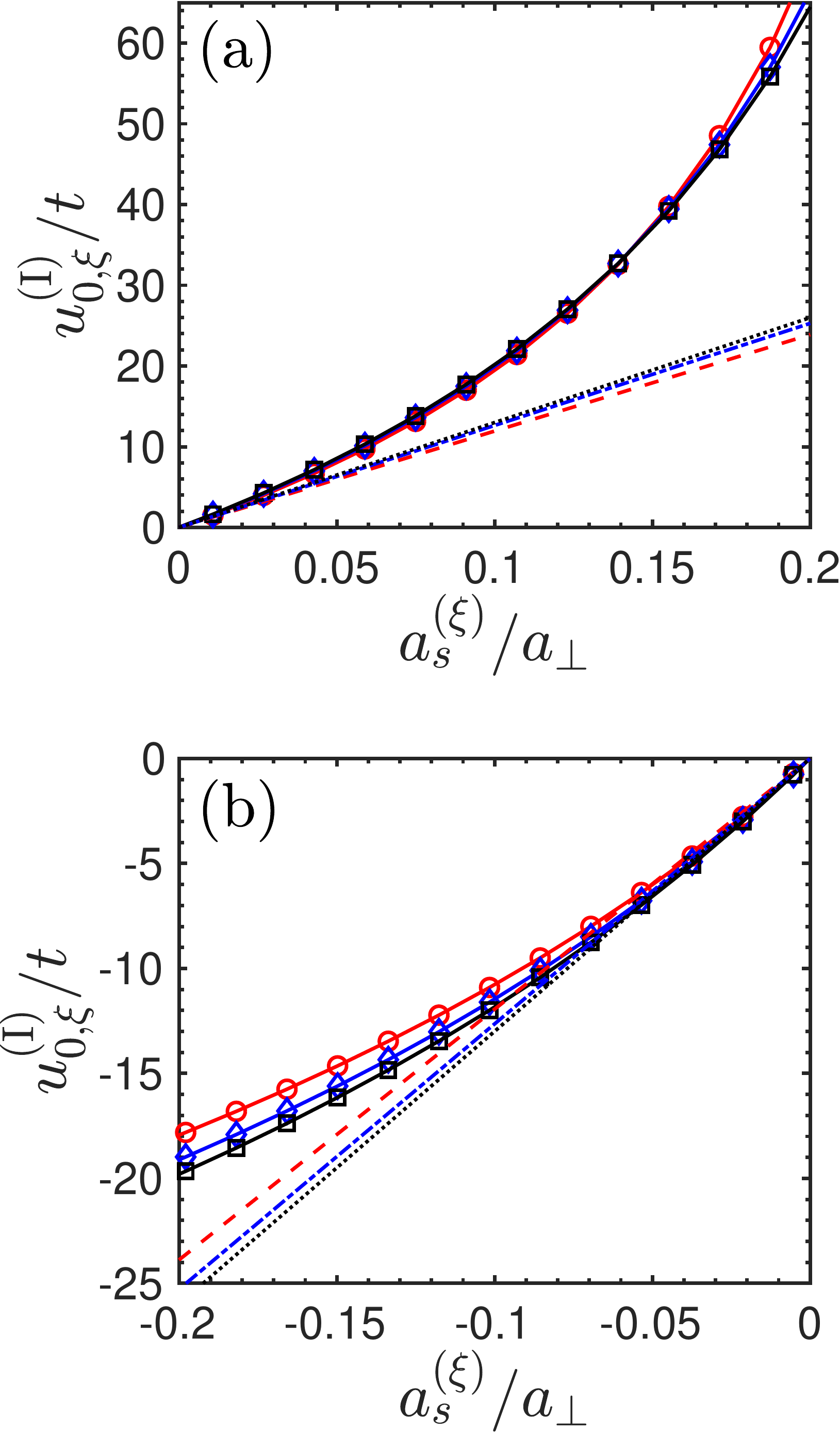} 
\caption{(color online) Comparison between the exact result of the interaction parameter $u_{0,\xi}^{\rm (I)}$ given by Eq. (\ref{u0iii})
(circles, diamonds, and squares connected by solid lines)
 and the value $u_{\rm project}$ given by the projection approximation (\ref{app}) 
 (dashed, dashed-dotted and dotted lines)
 for systems with $a_{s}^{(\xi)}>0$ {\bf (a)} and $a_{s}^{(\xi)}<0$ {\bf (b)} regime. Here we consider the cases with $a_{z}/a_{\perp}= 0.9$, $a_z/l_0=0.18$ (red circle and dashed line), $a_{z}/a_{\perp}= 0.8$, $a_z/l_0=0.15$ (blue diamond and dashed-dotted line), $a_{z}/a_{\perp}= 0.7$, $a_z/l_0=0.14$ (black square and dotted line), and $s_{g}=5$.
\label{projection}}
\end{figure}

Qualitatively speaking, this projection approximation is applicable when $a_s^{(\xi)}$ is ``much smaller" than all the characteristic lengths $\{a_z,a_\perp,l_0\}$ of the trapping potentials and the optical lattice. Here our question is, quantitatively speaking, for how small is $a_s^{(\xi)}$ we can use this approximation?

To answer this question, in Fig.~\ref{projection} we compare the exact value of $u_{0,\xi}^{\rm (I)}$ given by Eq. (\ref{u0iii}) with the approximated value $u_{\rm project}$ for a typical case with $a_{z}/a_{\perp}= 0.8$, $a_z/l_0= 0.15$, $s_{g}=5$. It is shown that in the limit of $a_{s}^{(\xi)}\to0$, the projection approximation works well. However, when the system derivates from the non-interacting point, the approximation fails very soon and the exact value rapidly becomes larger than $u_{\rm project}$, especially for the cases with $a_{s}^{(\xi)}>0$. To be specific,
the relative error of the projection approximation is already as large as $70\%$ when 
 $a_{s}^{(\xi)}/a_{\perp}=0.1$, and is about $200\%$ for $a_{s}^{(\xi)}/a_{\perp}=0.2$.
In addition, for $a_{s}^{(\xi)}<0$, when $a_{s}^{(\xi)}/a_{\perp}=-0.2$ the relative error of the projection approximation is also about $40\%$. Therefore, 
for our system, the approximation does not work even when the 3D scattering length is of the order of 10\% of the characteristic lengths of the confinement, and in these cases, one needs to determine the $u$-parameters via 
 the exact calculation for the inter-atomic low-energy scattering amplitudes or careful experimental calibrations.

\subsection{Spin-exchange interaction of ultracold $^{171}{\rm Yb}$ or $^{173}{\rm Yb}$ atoms}
Now we implement our approach to the $^{171}$Yb and $^{173}$Yb atomic gases and investigate the spin-exchange interaction in these two systems. Here we consider the setup of the experiment of Ref. \cite{kondo_exp}. 
In this experiment the transverse confinement potential is realized by a 2D optical lattice with magic wave lengths $\lambda_\perp=759$nm, while both the axial lattice for the $g$-atom and the axial trapping for the $e$-atom are realized via a 1D optical lattice with wave length $\lambda_z=670$nm, respectively. The explicit expression for the total transverse potential $U_{\rm lattice}^{(\perp)}$ for the two atoms and the axial potential $U_{\rm lattice}^{(z, j)}$ ($j=e,g$) for the $j$-atom are given by
\begin{eqnarray}
U_{\rm lattice}^{(\perp)}&=&s_{\perp}E_R^{(\perp)}\sum_{j={\rm e,g}}\left[\sin^2\left(\frac{2\pi x_j}{\lambda_\perp}\right)+\sin^2\left(\frac{2\pi y_j}{\lambda_\perp}\right)\right];\nonumber\\
\label{vlattice}\\
U_{\rm lattice}^{(z,e)}&=&s_{e}E_R\sin^2(k_0z_e);\label{vze}\\
U_{\rm lattice}^{(z,g)}&=&s_{g}E_R\sin^2(k_0z_g),\label{vzg}
\end{eqnarray}
where $x_j$ and $y_j$ ($j=e,g$) are the $x$- and $y$- coordinates of the $j$-atom, respectively, $E_R^{(\perp)}=2\hbar^{2}\pi^{2}/(m\lambda_{\perp}^{2})$, and $E_R=\hbar^{2}k_{0}^{2}/2m$ as defined before, and the dimensionless parameters $s_\perp$, $s_g$ and $s_e$ are the depths of these lattice potentials. 

\begin{figure}[t]
\centering
\includegraphics[width=0.45\textwidth]{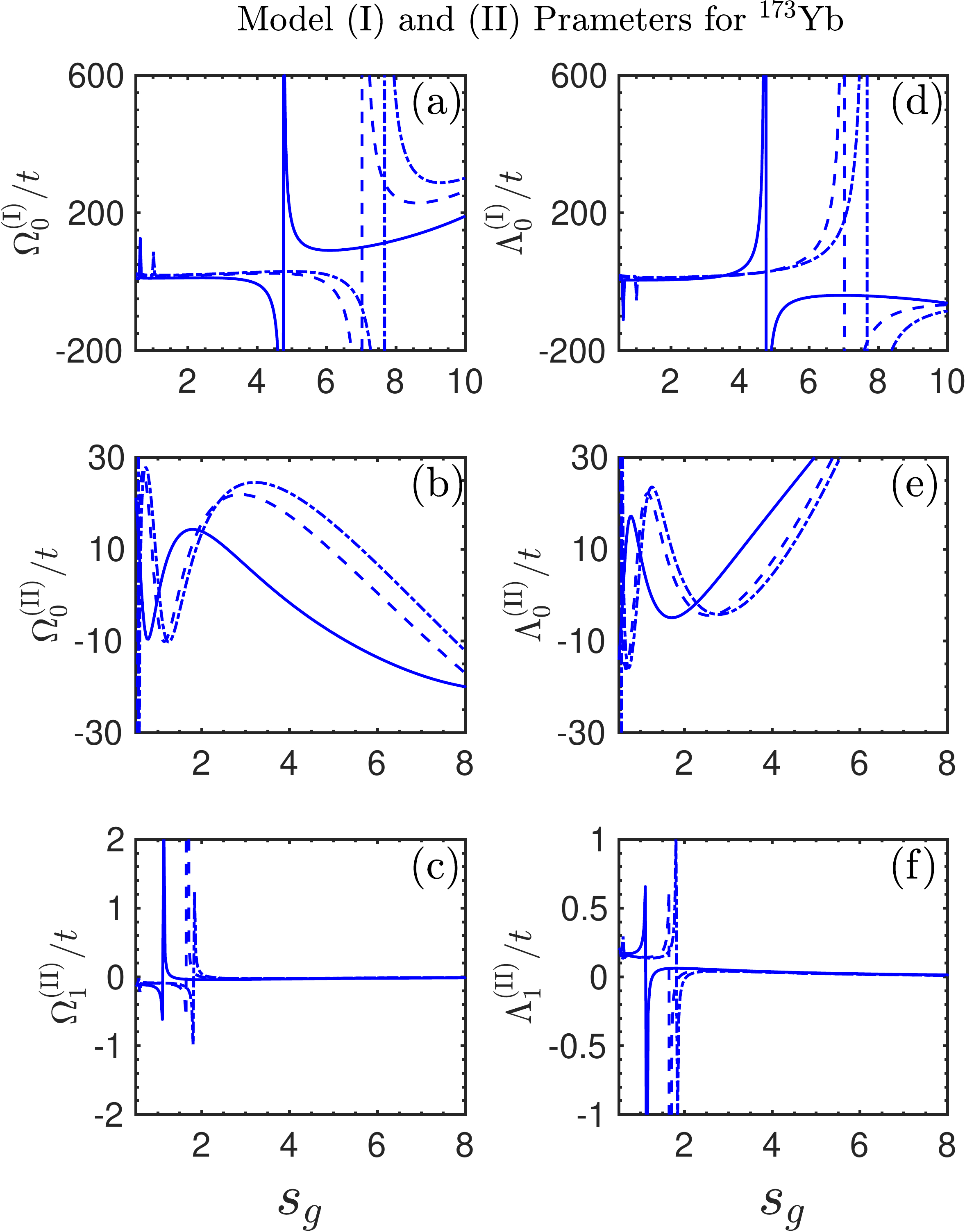}
\caption{ (color online) Spin-exchange interaction parameters $\Omega_{0}^{(\rm I)}$ { (a)}, $\Omega_{0,1}^{(\rm II)}$ { (b, c)} and spin-independent interaction $\Lambda_{0}^{(\rm I)}$ { (d)}, $\Lambda_{0,1}^{(\rm II)}$ { (e, f)} of the tight-binding models $H_{\rm MB}^{\rm (I)}$ and $H_{\rm MB}^{\rm (II)}$ for $^{173}$Yb atoms. Here we consider the case with the transverse confinement intensity $s_{\perp}=20$ (solid lines), $s_{\perp}=35$ (dashed lines) and $s_{\perp}=40$ (dash-dotted lines). Other parameters are given in Sec. IV. D.
\label{Yb173}}
\end{figure}

\begin{figure}[t]
\centering
\includegraphics[width=0.45\textwidth]{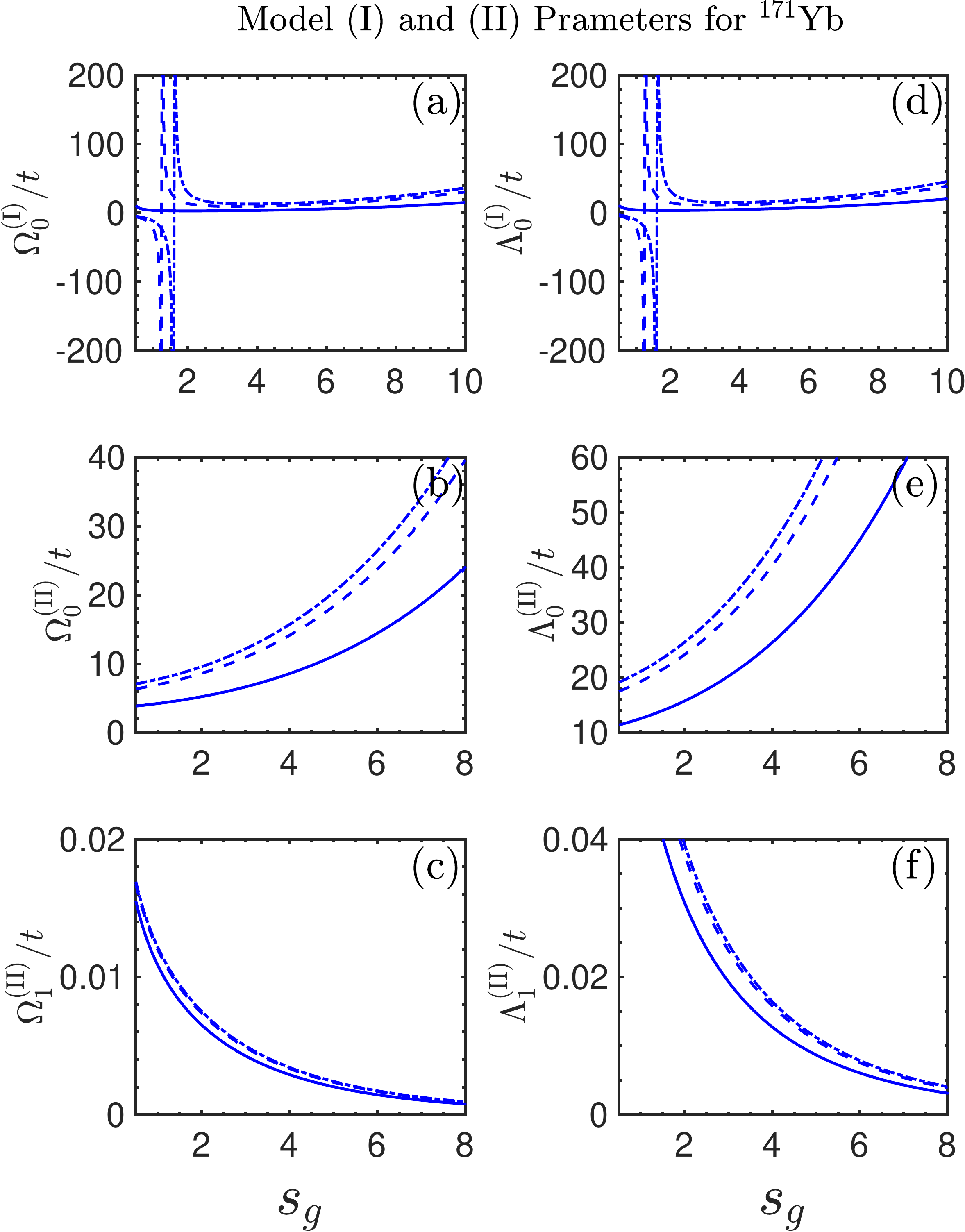}
\caption{ (color online)Spin-exchange interaction parameters $\Omega_{0}^{(\rm I)}$ { (a)}, $\Omega_{0,1}^{(\rm II)}$ { (b, c)} and spin-independent interaction $\Lambda_{0}^{(\rm I)}$ { (d)}, $\Lambda_{0,1}^{(\rm II)}$ { (e, f)} of the tight-binding models $H_{\rm MB}^{\rm (I)}$ and $H_{\rm MB}^{\rm (II)}$ for $^{171}$Yb atoms. Here we consider the case with the transverse confinement intensity $s_{\perp}=20$ (solid lines), $s_{\perp}=35$ (dashed lines) and $s_{\perp}=40$ (dash-dotted lines). Other parameters are given in Sec. IV. D.
\label{Yb171}}
\end{figure}

In our calculation we take the exact form of $U_{\rm lattice}^{(z,g)}$. In addition, as mentioned before, by expanding $U_{\rm lattice}^{(z,e)}$ and $U_{\rm lattice}^{(\perp)}$ around the minimum points we can obtain the harmonic trapping potentials shown in Eqs. (\ref{hperp}) and (\ref{hze}). Accordingly, for this system the characteristic lengths $a_z$, $a_\perp$ and the axial wave vector $k_0$ for the $g$-atom are given by
\begin{eqnarray}
a_\perp=\frac{\lambda_{\perp}}{\sqrt{2}\pi s_{\perp}^{1/4}};\
a_z=\frac{\lambda_{z}}{2\pi s_{e}^{1/4}};\ k_0=\frac{2\pi}{\lambda_z},
\end{eqnarray}
respectively. Moreover, as mentioned before, for $^{173}$Yb, the 3D scattering lengths are $a_s^{(+)}\approx1878a_{0}$ and $a_s^{(-)}\approx216a_{0}$ \cite{Yb173as,Fallani-PRX}, while for $^{171}$Yb we have $a_s^{(+)}\approx225a_{0}$ and $a_s^{(-)}\approx355a_{0}$ \cite{Yb171}. 

For our system, since the axial potentials for the $e$-atom and $g$-atom are created by the same laser beam, the intensities $s_{e}$ and $s_{g}$ for these potentials are related with each other. For the laser beam with wavelength $670$nm, we have $s_e\approx 3.3s_g$ for both $^{173}{\rm Yb}$ and $^{171}{\rm Yb}$ atoms, because the electronic structures of these two isotopes are quite similar. Thus, in the experiments, the independent control parameters are $s_\perp$ and $s_g$, which can be tuned via the intensities of the laser beams.


In Fig.~\ref{Yb173} and Fig.~\ref{Yb171} we present the  interaction strengths $\Omega_{0}^{\rm (I)}$, $\Lambda_{0}^{\rm (I)}$ of the tight-binding model $H_{\rm MB}^{\rm (I)}$ in Eq. (\ref{hmb1}) and $\Omega_{0,1}^{\rm (II)}$, $\Lambda_{0,1}^{\rm (II)}$ of the tight-binding model $H_{\rm MB}^{\rm (II)}$ in Eq. (\ref{hmb2}), for $^{173}$Yb atoms and $^{171}$Yb atoms respectively. According to Eq.(\ref{Omega}), these parameters are simple functions of the $u$-parameters.
 It is shown that the effective interaction can be efficiently controlled by the laser parameter $s_g$. On the other hand, Fig.~\ref{Yb173} and Fig.~\ref{Yb171} also shows that for the same case, the parameters $\eta_{0}^{\rm (I)}$ and $\eta_{0}^{\rm (II)}$ ($\eta=\Omega,\Lambda$) of the two tight-binding models can take quite different values, even if the system is not under an odd-wave CIR. This phenomenon can be explained as follows. According to Eq. (\ref{u0iii}) and Eq. (\ref{hubbardU0}), the interaction parameters $u_{0,\xi}^{\rm (I)}$ and $u_{0,\xi}^{\rm (II)}$ ($\xi=+,-$) take different values for the same case, even when the odd-wave scattering is weak so that the corresponding scattering length $a_{\rm o}^{(\xi)}$ is very small. For instance, when $a_{\rm e}^{(\xi)}=a_{\rm o}^{(\xi)}$, we have $u_{0,\xi}^{\rm (II)}=\infty$ while $u_{0,\xi}^{\rm (I)}$ is still finite. Furthermore, since $\eta_{0}^{\rm (I, II)}$ ($\eta=\Omega,\Lambda$) are simple functions of $u_{0,\pm}^{\rm (I, II)}$, the values of $\eta_{0}^{\rm (I)}$ and $\eta_{0}^{\rm (II)}$ can also be very different. Nevertheless, as shown in Fig.~\ref{fk}, when the odd-wave scattering can be neglected both of the the two tight-binding models can reproduce the same correct low-energy even-wave scattering amplitudes. That is because the even-wave scattering amplitude of $H_{\rm 1D}^{\rm (I)}$ (and $H_{\rm MB}^{\rm (I)}$) only depends on $u_{0,\xi}^{\rm (I)}$, while the one of $H_{\rm 1D}^{\rm (II)}$ (and $H_{\rm MB}^{\rm (II)}$) depends on both $u_{0,\xi}^{\rm (II)}$ and $u_{1,\xi}^{\rm (II)}$.

 \section{ Finite-momentum effect in the experiment of Ref. \cite{kondo_exp}}
 
 At the end of this work, we re-visit the experimental results of Ref. \cite{kondo_exp}. As introduced in Sec. IV.D, in this experiment the ultracold quasi-(1+0)D $^{173}{\rm Yb}$ atoms are trapped in the quasi-(1+0)D confinement. 
The localized $e$-atom and the moving $g$-atoms are initially prepared in the nuclear-spin states $|\downarrow\rangle_e$ and $|\uparrow\rangle_g$, respectively. 
After a finite holding time, the nuclear spin of some $e$-atoms are flipped to $|\uparrow\rangle_e$ by the inter-atomic spin-exchange collision. The number $N_{e\uparrow}$ of these spin-flipped $e$-atoms is measured.

We study the spin-exchange collision rate $R_{\rm se}$ which depends on incident quasi-momentum $k$ of the $g$-atom. Using this result we will further investigate the finite-momentum effect of this experiment.
We first show the definition of $R_{\rm se}$. According to the above description, the incident wave function of the spin-exchanging collision in the experiment can be expressed as
\begin{eqnarray}
|\Phi_{\rm in}({\bm \rho},z_{g},z_{e})\rangle&=&\chi_0({\bm \rho})\phi_{0}(z_{e})\psi(z_{g})|\uparrow\rangle_g|\downarrow\rangle_e,\nonumber\\
&=&\frac{1}{\sqrt{2}}\chi_0({\bm \rho})\phi_{0}(z_{e})\psi(z_{g})\left(|+\rangle+|-,0\rangle\right),\nonumber\\
\label{ss1}
\end{eqnarray}
where the functions $\chi_0({\bm \rho})$, $\phi_{0}(z_{e})$ and $\psi(z_{g})$ are defined in Sec. IV. A, and the spin singlet state $|+\rangle$ and triplet state $|-,0\rangle$ are defined in Eq. (\ref{p}) and Eq. (\ref{m0}), respectively. Furthermore, the scattering process is governed by the Hamiltonian $H$ of Eq. (\ref{h}), which is diagonal in the singlet-triplet basis. Thus, the scattering wave function corresponding to $|\Phi_{\rm in}({\bm \rho},z_{g},z_{e})\rangle$ can be expressed as
\begin{eqnarray}
&&|\Phi^{(+)}({\bm \rho},z_{g},z_{e})\rangle\nonumber\\
&=&\frac{1}{\sqrt{2}}\left[\Psi_{+}(\bm{\rho},z_{e},z_{g})|+\rangle+\Psi_{-}(\bm{\rho},z_{e},z_{g})|-,0\rangle\right].
\end{eqnarray}
Here the function $ \Psi_{\xi}(\bm{\rho},z_{e},z_{g})$ ($\xi=+,-$) is the one calculated in Sec. IV. A. Using the out-going boundary condition (\ref{psi2}) of $ \Psi_{\xi}(\bm{\rho},z_{e},z_{g})$, we can obtain the behavior of $|\Phi^{(+)}({\bm \rho},z_{g},z_{e})\rangle$ in the limit that the two atoms are far away from each other 
(i.e., the limit $|z_{g}|\rightarrow\infty$):

\begin{widetext}
\begin{eqnarray}
|\Phi^{(+)}({\bm \rho},z_{e},|z_{g}|\rightarrow\infty)\rangle&=&
|\Phi_{\rm in}({\bm \rho},z_{g},z_{e})\rangle
+ \chi_0({\bm \rho})\phi_{0}(z_{e})\left[
 Q_{\rm ela}^{\rm (e)}(k) e^{ik|z_{g}|}{\cal F}(|z_{g}|)+
 Q_{\rm ela}^{\rm (o)}(k){\rm sign}(z_g)e^{ik|z_{g}|}{\cal F}(|z_{g}|)
 \right]|\uparrow\rangle_g|\downarrow\rangle_e\nonumber\\
&& + \chi_0({\bm \rho})\phi_{0}(z_{e})\left[
 Q_{\rm se}^{\rm (e)}(k) e^{ik|z_{g}|}{\cal F}(|z_{g}|)+
 Q_{\rm se}^{\rm (o)}(k){\rm sign}(z_g)e^{ik|z_{g}|}{\cal F}(|z_{g}|)
 \right]|\downarrow\rangle_g|\uparrow\rangle_e. \label{psi2aa}
\end{eqnarray}
\end{widetext}
Here we also used the relations (\ref{p}, \ref{m0}) between the basis $\{|+\rangle,|-,0\rangle\}$ and $\{|\uparrow\rangle_g|\downarrow\rangle_e,|\downarrow\rangle_g|\uparrow\rangle_e \}$. In Eq. (\ref{psi2aa}) $Q_{\rm ela}^{\rm (e/o)}$ are the even/odd wave scattering amplitudes for the elastic process where the nuclear-spin of the two atoms is not changed, and $Q_{\rm se}^{\rm (e/o)}$ are the even/odd wave scattering amplitudes for the spin-exchange collision. They can be expressed as ($s$=e,o)
\begin{eqnarray}
Q_{\rm ela}^{(s)}(k)=\frac{1}{2}\left[f^{(s)}_-(k)+f^{(s)}_+(k)\right];\\
Q_{\rm se}^{(s)}(k)=\frac{1}{2}\left[f^{(s)}_-(k)-f^{(s)}_+(k)\right],
\end{eqnarray}
where $f^{(s)}_\xi(k)$ ($s$=e,o, $\xi=+,-$) are the scattering amplitudes for the singlet and triplet channels, which are defined and calculated in Sec. IV. A. 
Furthermore, the spin-exchange collision rate $R_{\rm se}$ can be defined as
\begin{eqnarray}
R_{\rm se}(k)=|Q_{\rm se}^{\rm (e)}(k)|^2+|Q_{\rm se}^{\rm (o)}(k)|^2.\label{ss2}
\end{eqnarray}

\begin{figure}[t]
\centering
\includegraphics[width=0.46\textwidth]{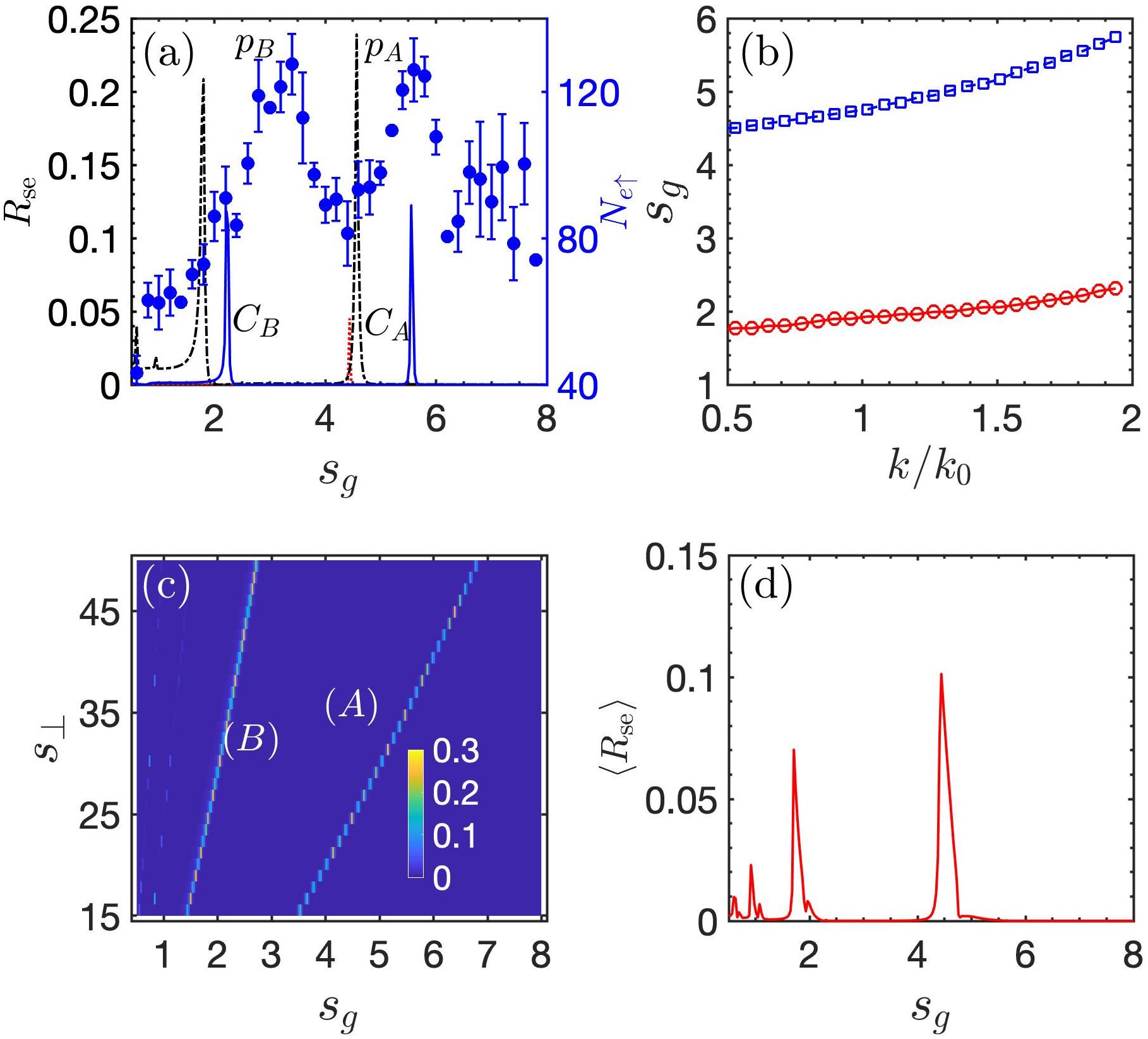} 
\caption{(color online) Spin-exchange collision rate $R_{\rm se}$ for the experimental system of Ref. \cite{kondo_exp}. {\bf (a)}: $R_{\rm se}(k)$ as a function of $s_g$, for $s_\perp=35$, with incident quasi-momentum $k$ of the $g$-atom taking values $k=0.01k_{0}$(red dotted), $k=0.6k_{0}$(black dash-dotted) and $k=1.8k_{0}$(green solid). Here we also show the experimental datas reprinted for the observed spin-flipped atom number $N_{e\uparrow}$ from Ref. \cite{kondo_exp}. {\bf (b)}: Locations of two peaks of $R_{\rm se}$ (corresponding to $C_{A}$ and $C_{B}$ for $k=1.8k_0$) in (a), for various values of $k$. {\bf (c)}: $R_{\rm se}(k)$ for $k=1.8k_0$ in the parameter region $s_{g}\in[0.5,8]$ and $s_{\perp}\in[15,50]$. {\bf (d)}: Averaged spin-exchange collision rate $\langle R_{\rm se}\rangle$ as function of $s_{g}$, for the case of (a). Other details of the calculation of $\langle R_{\rm se}\rangle$ are introduced in the last paragraph in Sec. V.} \label{com-exp}
\end{figure}

In Fig.~\ref{com-exp}(a) we show the spin-exchange collision rate $R_{\rm se}$ of $^{173}$Yb atoms as a function of the axial lattice intensity $s_g$, 
for the experimental case with the transverse confinement intensity $s_{\perp}=35$ (the definitions of $s_{g,\perp}$ are given in Sec. IV. D), for various incident quasi-momentum $k$ of the $g$-atom. It is clearly shown that the behavior of $R_{\rm se}$ strongly depends on $k$. Specifically, when $k$ is increased the peaks of $R_{\rm se}$ can be significantly shifted to larger $s_{g}$. This is also shown in  
 Fig.~\ref{com-exp}(b) where we illustrate the locations of two peaks in Fig.~\ref{com-exp}(a) as functions of $k$. These results indicate that the finite-momentum effect is very important for the system in the experiment of Ref. \cite{kondo_exp}.

 We further compare our results with the experimental measurement for the spin-flipped $e$-atom number $N_{e\uparrow}$, which are given in Ref.~\cite{kondo_exp}. As shown in Fig.~\ref{com-exp}(a), when the incident quasi-momentum $k$ takes the value $k=1.8k_{0}$ \cite{momentum}, the location of one peak ($C_A$) of $R_{\rm se}$ is quantitatively consistent with the peak $p_{A}$ of $N_{e\uparrow}$, and the other peak ($C_B$) of $R_{\rm se}$ is also close to another peak $p_{B}$ of $N_{e\uparrow}$. The difference between the positions of $C_B$ and $p_B$ may originate from the anharmonicity of the axial trapping potential of the $e$-atom.

Moreover, in Fig.~\ref{com-exp}(c) we present the spin-exchange collision rate $R_{\rm se}$ for the case with $k=1.8k_{0}$ in the parameter region
$s_{g}\in[0.5,8]$ and $s_{\perp}\in[15,50]$. Two branches (A) and (B) 
with large $R_{\rm se}$
are clearly shown. Our calculations show that they are induced by the even- and odd-wave CIR, respectively. 
Comparing our results with Fig.~3(a) of Ref. \cite{kondo_exp} where the observed CIR branches are illustrated, we find that the location of the even-wave CIR branch (A) is quantitatively consistent the one observed in the whole parameter region. The location of the odd-wave CIR branch (B) is also close to another observed branch, and the difference of their locations is possibly due to the anharmonicity of the longitudinal confinement potential for the $e$-atom, as discussed above. Besides, Fig.~\ref{com-exp}(c) also shows some CIR branches around $s_{g}=1$, which are too narrow to be observed in the experiment.

As shown above, the positions of the CIR branches given by our theoretical calculation and the experimental measurements in Ref.~\cite{kondo_exp} can be quantitatively consistent with each other only when the incident quasi-momentum of the $g$-atom is $k=1.8k_0$. Thus, it is possible that the quasi-momentum of the $g$-atoms in the experiment of Ref.~\cite{kondo_exp} is as high as $1.8k_0$, which is already in the second Brillouin zone. If that is the case, 
in the experiment most of the $g$-atoms are in the second band of the optical lattice. As a result, further cooling may be necessary to realize the single-band tight-binding model. 
Nevertheless, there is also another possibility. Our calculation is done for the spin-exchange rate of a single collision between one $g$-atom and the $e$-atom, while the experiment was done for many $g$-atoms that can collide with the $e$-atom for many times. The many-body and multi-collision effects, which are beyond our theoretical treatment,  may make the value of $k$ given by our theoretical fitting to be higher than its actual value \cite{anharmonicity}. If that is true, the realistic
$g$-atom quasi-momentum in the experiment may not be so high.

Fig.~\ref{com-exp}(a) also shows that the observed peaks $p_{A,B}$ are broader than the ones $C_{A,B}$ from our calculation. One can also see this by comparing Fig.~3(a) of Ref. \cite{kondo_exp} and our Fig.~\ref{com-exp}(c). This phenomenon may originate from the fact that in the experiment the $g$-atom quasi-momentum $k$ does not take a certain value, but satisfies a probability distribution $p(k)$. As a result, it is the average spin-exchange collision rate 
\begin{align}
\langle R_{\rm se}\rangle=\int R_{\rm se}(k)p(k)dk
\end{align}
that corresponds to the experimental observation. The peak of the spin-exchange collision rate can be broadened by this averaging. 
To illustrate this broaden effect, we calculate $\langle R_{\rm se}\rangle$ for the case of Fig.~\ref{com-exp}(a) with $p(k)=e^{-{\cal E}(k)/{\cal E}(k_0)}/Z$. Here $Z$ is the normalized factor and ${\cal E}(k)$ is the dispersive relation of the $g$-atoms in the lowest band of the axial lattice. We show our results in Fig.~\ref{com-exp}(d). By comparing Fig.~\ref{com-exp}(d) and Fig.~\ref{com-exp}(a), one could find that 
the peaks of the spin-exchange collision rate is indeed broadened by the thermal distribution of the $g$-atom quasi-momentum.


\section{summary}
In this work, we show that the ultracold gases of alkaline-earth atoms 
in electronic $e$- and $g$-states, which are trapped
 in the quasi (1+0)D confinement with an axial optical lattice can be described by two tight-binding Kondo models (I) and (II) for two different cases. We further derive the interaction parameters for these two models by exactly solving exactly the two-atom scattering problem of this system.
The comparison between the exact results and the ones given by the simple projection approximation shows that for our system this approximation fails even when the 3D scattering length is only
of the order of 10\% of the characteristic lengths of the confinement and optical lattice potentials. 
By implementing our theory to ultracold $^{173}$Yb and $^{171}$Yb atoms, we
further illustrate the control effects of the confinement and optical lattice on the interaction parameters of these two models. Moreover,
using the exact solution of the two-body scattering problem, we also derive the spin-exchanging rate of $^{173}$Yb atoms in the recent experiment of Ref. \cite{kondo_exp}, and find that for this experiment the effect of the finite-momentum of the $g$-atoms is very significant.

\begin{acknowledgments}
This work is supported by the National Key R$\&$D Program of China (Grant No. 2018YFA0306502 (PZ), 2018YFA0307601 (RZ)), NSFC Grant No. 11804268 (RZ), 11434011(PZ), 11674393(PZ), as well as the Research Funds of Renmin University of China
under Grant No. 16XNLQ03(PZ).
\end{acknowledgments}

\appendix

\begin{widetext}
\section{Scattering Amplitude Calculation}
\label{sec:green}
In this Appendix, we show the detailed calculation of scattering amplitude corresponding to the Hamiltonian $H_\xi$ in Eq. (\ref{hxi}). 

\subsection{1D Lattice Green's function}
For future convenience, we start with the derivation of the Green's function for a single atom in an 1D optical lattice. The Hamiltonian for such system can be written as
\begin{align}
H_{{\rm 1D}}=-\frac{\hbar^{2}}{2m}\frac{d^{2}}{dz^{2}}+sE_{R}\sin^{2}\left(k_{0}z\right),
\end{align}
where $m$ is the single atom mass, $E_{R}=\hbar^{2}k_{0}^{2}/2m$ is the recoil energy with $k_{0}$ being the wave vector, and $s$ denotes lattice depth. The corresponding 1D Green's function $G_{\rm 1D}({\cal E};z,z')$ satisfies
\begin{align}
\left({\cal E}-H_{\rm 1D}\right)G^{\rm (1DL)}({\cal E};z,z')=\delta(z-z'),
\end{align}
as well as the out-going boundary condition in the limit $|z|\rightarrow\infty$.
Thus, when $z\neq z'$, the 1D Green's function satisfies the Mathieu equation
\begin{align}
\label{mathieu}
\left[\frac{d^{2}}{dz^{2}}+\left(a-2q\cos(2k_0z)\right)\right]G^{\rm (1DL)}({\cal E};z,z')=0,
\end{align}
with $a=2m{\cal E}/(\hbar^{2}k_{0}^{2})-s/2$, $q=-s/4$. For $z\neq z'$, this equation has two linearly-independent solutions, i.e., $e^{ikz}{\cal F}(z)$ and $e^{-i z}{\cal F}(-z)$, where 
the function ${\cal F}(z_g)$ is defined as ${\cal F}(z_g)={\cal M}(\pi z_g/l_0)$, with ${\cal M}(s)$ being the Mathieu function of the first kind with period $\pi$.
In Eq. (\ref{mathieu}), $k$ is the characteristic exponent of this Mathieu equation. It depends on ${\cal E}$ and satisfies $k>0$.

Furthermore, 
the connection conditions of $G^{\rm (1DL)}(E;z,z')$ at $z=z'$ can be expressed as
\begin{align}
&G^{\rm (1DL)}(E;z^{\prime}+0^{+},z^{\prime}) = G^{\rm (1DL)}(E;z^{\prime}-0^{+},z^{\prime});\label{cc1}\\
&\left.\frac{d}{dz}G^{\rm (1DL)}(E;z,z^{\prime})\right|_{z=z^{\prime}-0^{+}}^{z=z^{\prime}+0^{+}} = \frac{2m}{\hbar^{2}}.\label{cc2}
\end{align}
Solving Eqs. (\ref{mathieu},\ref{cc1},\ref{cc2}) together with the out-going condition, one immediately finds that $G_{\rm 1D}(E;z,z')$ can be written as
\begin{align}
\label{latticeG}
G^{\rm (1DL)}(E;z,z')=\left\{ \begin{aligned}
&\frac{2m}{\hbar^{2}}\frac{{\cal F}\left(z\right){\cal F}\left(-z'\right)}{{\cal W}(z')}e^{ik(z-z')},\ {\rm for}\ z>z^{\prime}\\
&\frac{2m}{\hbar^{2}}\frac{{\cal F}\left(-z\right){\cal F}\left(z'\right)}{{\cal W}(z')}e^{ik(z'-z)},\ {\rm for}\ z<z^{\prime}
\end{aligned}
\right.,
\end{align}
with ${\cal W}(z')=\left[{\cal F}\left(-z'\right){\cal F}'\left(z\right)+{\cal F}\left(z'\right){\cal F}'\left(-z'\right)+2ik{\cal F}\left(z'\right){\cal F}\left(-z'\right)\right]$. 

In the case of $s\to0$, ${\cal F}\left(z\right)$ becomes a constant and the quasi-momentum reduces to the momentum $k$. As a result, the Green's function in Eq.(\ref{latticeG}) reduces to 
\begin{eqnarray}
G^{\rm (1DF)}(E;z,z')\equiv\frac{m}{\hbar}\frac{e^{i\frac{\sqrt{2mE}}{\hbar}|z-z'|}}{i\sqrt{2mE}},\label{g1df}
\end{eqnarray}
with $\sqrt{E}=i\sqrt{|E|}$ for $E<0$, which is just the Green's function in free space.

\subsection{Lippmann-Schwinger Equation and Scattering Amplitude}

Now we return to our quasi-(1+0)D scattering problem governed by the Hamiltonian $H_\xi$ in Eq. (\ref{hxi}). Using the expression of the Huang-Yang pseudo potential $U_\xi$, we find that the scattering state $\Psi_{\xi}(\bm{\rho},z_{e},z_{g})$ corresponding to the incident wave function $\Psi_{\rm in}({\bm \rho},z_{g},z_{e})$ introduced in Eq. (\ref{psiin}) satisfies the Lippman-Schwinger type equation
\begin{align}
\Psi_{\xi}(\bm{\rho},z_{e},z_{g})=\Psi_{\rm in}({\bm \rho},z,z)+
\frac{2\pi\hbar^2a_s^{(\xi)}}{\mu}\int dz'{\cal G}({\bm \rho},z_{g},z_{e};{\bf 0},z',z')\eta(z'),
\label{lse}
\end{align}
where the function $\eta(z)$ is defined as
\begin{align}
\eta(z)=\left.\frac{\partial}{\partial b}\left[b\Psi_\xi\left({\bf 0},z+b/2,z-b/2\right)\right]\right|_{b\to0^+},\label{eta}
\end{align}
and the Green's function ${\cal G}$ associated with the free Hamiltonian in Eq.(\ref{H0}) can be expressed as (with the Dirac notations)
\begin{align}
{\cal G}({\bm \rho},z_{g},z_{e};{\bm \rho}',z_{g}',z_{e}')=\langle{\bm \rho},z_{g},z_{e}|\frac{1}{E+i0^{+}-H_{0}}|{\bm \rho}',z_{g}',z_{e}'\rangle.\label{G0}
\end{align}
Here $|{\bm \rho},z_{g},z_{e}\rangle$ is the eigen-state of the inter-atomic relative transverse coordinate and the axial coordinates of the two atoms. 
Furthermore, for ${\bm \rho}={\bm \rho}'={\bf 0}$, the Green's function can be simplified as
\begin{align}
{\cal G}({\bf 0},z_{g},z_{e};{\bf 0},z_{e}',z_{g}')&=\sum_{n_{\perp}=0,2,4,...}\frac{\mu\omega_{\perp}}{\pi\hbar}G_{n_{\perp}}(z_{g},z_{e};z_{e}',z_{g}'),
\label{g02}
\end{align}
with $G_{n_{\perp}}(z_{g},z_{e};z_{e}',z_{g}')$ being define as
\begin{align}
G_{n_{\perp}}(z_{g},z_{e};z_{e}',z_{g}') = & \langle z_{g},z_{e}|\frac{1}{E-(n_{\perp}+1)\hbar\omega_{\perp}+i0^{+}-\left[T_{z}+s_{z,g}E_{R}\sin^{2}\left(k_{0}z_{g}\right)+\frac{1}{2}m_{e}\omega_{z}^{2}z_{e}^{2}\right]}|z_{g}',z_{e}'\rangle\nonumber\\
=&\sum_{n_{z}=0,1,2,...}G^{\rm (1DL)}\left({\cal E}_{n_\perp,n_z};z_{g},z_{g}'\right)\phi_{n_{z}}(z_{e})\phi^{*}_{n_{z}}(z_{e}^{\prime}),\label{ggk}
\end{align}
where
\begin{eqnarray}
{\cal E}_{n_\perp,n_z}=E-(n_{\perp}+1)\hbar\omega_{\perp}-(n_{z}+\frac{1}{2})\hbar\omega_{z}.\label{ecal}
\end{eqnarray}
Here $G^{\rm (1DL)}\left(E;z_{g},z_{g}'\right)$ is the free Green's function given in the above subsection, and $\phi_{n_{z}}(z_{e})$ is the $n_z$-th eigen-state of $H_z^{(e)}$ defined in Eq. (\ref{hze}). In the derivation of Eq. (\ref{g02}) we have used the fact that 
$\left|\chi_{n_{\perp}}(\rho=0)\right|^{2}=\mu\omega_{\perp}/\hbar\pi$, where $\chi_{n_{\perp}}({\bm \rho})$ is the eigen-state of $H_\perp$ defined in Eq. (\ref{hperp}), with principle quantum number $n_\perp$ and zero axial angular momentum.

Substituting Eq. (\ref{g02}) and Eq. (\ref{ggk}) into Eq. (\ref{lse}), and using the expression (\ref{latticeG}) of $G^{\rm (1DL)}\left(E;z_{g},z_{g}'\right)$ and the boundary condition given in Eq.(\ref{newpsi}) of the main text, we find that the reflection and transmission amplitude defined in Eq.(\ref{newpsi}) can be expressed as 
\begin{align}
\label{reflection}
r_{\xi}=&\frac{2g}{\sqrt{\pi}a_{\perp}}\int dz'\frac{e^{ikz'}{\cal F}\left(z'\right)\phi^{*}_{0}(z')\eta(z')}{{\cal W}(z')},\\
\label{transmission}
t_{\xi}=&1+\frac{2g}{\sqrt{\pi}a_{\perp}}\int dz'\frac{e^{-ikz'}{\cal F}\left(-z'\right)\phi^{*}_{0}(z')\eta(z')}{{\cal W}(z')}
\end{align}
with $a_{\perp}=\sqrt{\hbar/\mu\omega_{\perp}}$. 
On the other hand, as shown in Sec. IV. A of our main text, 
$r_{\xi}$ and $t_{\xi}$ are related to the 
scattering amplitudes $f_{\xi}^{\rm (e/o)}$ via 
\begin{align}
f_{\xi}^{\rm (e)}=&\frac{1}{2}\left(r_{\xi}+t_{\xi}-1\right),\\
f_{\xi}^{\rm (o)}=&\frac{1}{2}\left(-r_{\xi}+t_{\xi}-1\right).\label{fff}
\end{align}

Therefore, if we can obtain the function $\eta(z)$, we can directly obtain the scattering amplitudes $f_{\xi}^{\rm (e/o)}$ via Eq. (\ref{transmission}-\ref{fff}).

\subsection{Calculation of $\eta(z)$}

Now we show how to calculate the function $\eta(z)$. Inserting Eq.(\ref{lse}) into Eq.(\ref{eta}), one immediately obtains the integral equation satisfied by $\eta(z)$,
\begin{align}
\eta(z)=&\Psi_{\rm in}({\bm 0},z,z)+\frac{2\pi\hbar^2a_s^{(\xi)}}{\mu}\left.\frac{\partial}{\partial b}\int dz'\left[b{\cal G}\left({\bf 0},z+b/2,z-b/2;{\bf 0},z',z'\right)\eta(z')\right]\right|_{b\to0^+}.\label{eta2}
\end{align}
For the convenience of the following calculation, we further express the Green's function ${\cal G}({\bf 0},z_{g},z_{e};{\bf 0},z',z')$ as
\begin{align}
\label{sub2}
 {\cal G}({\bf 0},z_{g},z_{e};{\bf 0},z',z') = \tilde{\cal G}({\bf 0},z_{g},z_{e};{\bf 0},z',z')+{\cal G}'(z_{g},z_{e},z'),
\end{align}
where $\tilde{\cal G}({\bm \rho},z_{g},z_{e};{\bm \rho},z_{g}',z_{e}')]$  
is defined as
\begin{align}
\tilde{\cal G}({\bm \rho},z_{g},z_{e};{\bm \rho}',z_{g}',z_{e}')=\langle{\bm \rho},z_{g},z_{e}|\frac{1}{E'+i0^{+}-[H_{\perp}-\frac{\hbar^{2}}{2m}\frac{d^{2}}{dz_{e}^{2}}+\frac{1}{2}m_{e}\omega_{z}^{2}z_{e}^{2}]}|{\bm \rho}',z_{g}',z_{e}'\rangle,\label{Gt}
\end{align}
with $E'=(\frac{\hbar^2k^2}{2m}+\hbar\omega_\perp+\frac{\hbar}2\omega_z)$. It is the free Green function of the Hamiltonian for the systems without the axial optical lattice, and ${\cal G}'(z_{g},z_{e},z')$ is given by
\begin{eqnarray}
{\cal G}'(z_{g},z_{e},z')&=&{\cal G}({\bf 0},z_{g},z_{e};{\bf 0},z',z')-\tilde{\cal G}({\bf 0},z_{g},z_{e};{\bf 0},z',z')\nonumber\\
&=&\frac{\mu\omega_{\perp}}{\pi\hbar}\sum_{n_{z}=0,1,2,\cdots}\sum_{n_{\perp}=0,2,4,\cdots}\left[G^{\rm (1DL)}\left({\cal E}'_{n_\perp,n_z};z_g,z'\right)-G^{\rm (1DF)}\left({\cal E}'_{n_\perp,n_z};z_g,z'\right)\right]\phi_{n_{z}}(z_e)\phi^{*}_{n_{z}}(z^{\prime}),
\end{eqnarray}
with ${\cal E}'_{n_\perp,n_z}=\frac{\hbar^2k^2}{2m}-n_\perp\omega_\perp-n_z\omega_z$. Here the functions $G^{\rm (1DL)}$, $G^{\rm (1DF)}$ and the parameter ${\cal E}_{n_\perp,n_z}$ are defined in Eq. (\ref{latticeG}), Eq. (\ref{g1df}) and Eq. (\ref{ecal}), respectively.
We can submit Eq. (\ref{sub2}) into Eq. (\ref{eta2}), and re-write the integral equation for $\eta(z)$ in terms of the functions $\tilde{\cal G}({\bf 0},z_{g},z_{e};{\bf 0},z',z')$ and ${\cal G}'(z_{g},z_{e},z')$. Furthermore, we notice that in the limit $z_{g}\to z_{e}$ the behavior of the function ${\cal G}({\bf 0},z_{g},z_{e};{\bf 0},z',z')$ is irregular, while the behavior of ${\cal G}'(z_{g},z_{e},z')$ is regular. Fortunately, in our previous work \cite{renkondo2} for the quasi-(1+0)D system without the axial lattice, we have treated the irregularity of the Green's function ${\cal G}({\bf 0},z_{g},z_{e};{\bf 0},z',z')$. Here we can directly use the approach in that work. After straightforward calculation, we eventually obtain 
\begin{eqnarray}
\eta(z) &= & \Psi_{\rm in}({\bm 0},z,z)+\frac{2\pi\hbar^{2} a_{s}}{\mu}\left\{F_{1}(z)\eta(z)+\int dz'F_{2}(z,z^{\prime})\left[\eta(z')-\eta(z)\right]\right\}+2\hbar\omega_{\perp}a_{s}^{(\xi)}\int dz^{\prime}\left[{\cal G}'(z,z,z')+F_3(z,z')\right]\eta(z'),\nonumber\\ \label{final}
\end{eqnarray}
with
\begin{eqnarray}
F_{1}(z)&=& \frac{-\mu}{(2\pi\hbar^{2})^{3/2}\sqrt{2}}\int_{0}^{\infty}d\beta\left[\frac{\hbar\omega_{\perp}\sqrt{2\omega_{z}}\exp\left(\beta E^{\prime}-\frac{m\omega_{z}\left[\omega_{z}\hbar\beta+2\tanh(\hbar\omega_{z}\beta/2)\right]}{2\hbar[1+\omega_{z}\hbar\beta\coth(\hbar\omega_{z}\beta)]}z^{2}\right)}{\sinh(\hbar\beta\omega_{\perp})\sqrt{\omega_{z}\beta\cosh(\hbar\omega_{z}\beta)+\sinh(\hbar\omega_{z}\beta)/\hbar}}-\frac{1}{\beta^{3/2}}\right];
\label{f1}\\
F_{2}(z,z^{\prime})&=&-\int_{0}^{\infty}d\beta e^{\beta E^{\prime}}\frac{m\mu\omega_{\perp}}{(2\pi\hbar)^{2}\sinh(\hbar\omega_{\perp}\beta)}\sqrt{\frac{\omega_{z}}{\hbar\beta\sinh(\hbar\omega_{z}\beta)}} \exp\left[-\frac{m\omega_{z}[\left( z^{2}+z^{\prime2}\right) \cosh(\hbar\omega_{z}\beta)-2zz^{\prime}]}{2\hbar\sinh(\hbar\omega_{z}\beta)}-\frac{m(z-z^{\prime})^{2}}{2\hbar^{2}\beta}\right];\nonumber\\
\\
F_3(z;z')&=&\sum_{n_{z}=0}^{\infty}
G^{\rm (1DF)}\left({\cal E}'_{n_\perp=0,n_z};z,z'\right)
\phi_{n_{z}}(z)\phi_{n_{z}}^{*}(z'),\label{littleg0}
\end{eqnarray}
where $m$ is the single-atom mass.

Eq. (\ref{final}) is an integral equation for $\eta(z)$, with all singularities being removed. We numerically solve this equation and obtain the function of $\eta(z)$. Using this result we can directly calculate the scattering amplitudes $f_{\xi}^{\rm (e/o)}$ with the approach shown in the above subsection.

\section{Interaction Parameters of Tight-Binding Models}
\label{sec:tightbinding}
In this Appendix we prove Eq. (\ref{u0iii}-\ref{hubbardU1}) of our main text. 

We first prove Eq. (\ref{hubbardU0}) and Eq. (\ref{hubbardU1}).
The principle philosophy is that we expect the 1D tight-binding model
\begin{eqnarray}
h_\xi^{\rm (II)}\equiv T_{\rm 1D}+ u_{0,\xi}^{\rm (II)}|0\rangle_g\langle 0|+u_{1,\xi}^{\rm (II)}\sum_{i=\pm1}|i\rangle_g\langle i|,
\end{eqnarray}
with $\xi=+,-$ and $T_{\rm 1D}$ being given by Eq. (\ref{t1d}) of the main text, can reproduce the even- and odd-wave scattering lengths $a^{\rm (e/o)}_{\xi}$ of the exact Hamiltonian $H$. To this end, we need to solve the scattering problem corresponding to $h_\xi^{\rm (II)}$.
The scattering wave function of $h_\xi^{\rm (II)}$ can be written as
\begin{align}
\label{scatteringstate-lattice}
|k+\rangle=\sum_{n=-\infty}^{\infty}\psi_{k}^{(+)}(n)|n\rangle_g,
\end{align}
where $k\in[0,k_0]$ is the incident quasi-momentum, and $\psi_{k}^{(+)}(n)$ is the wave function defined on the optical lattice. 
Substituting scattering state given in Eq.(\ref{scatteringstate-lattice}) into the Shr\"odinger equation $h_\xi^{\rm (II)}|k+\rangle=\epsilon_{k}|k+\rangle$ with $\epsilon_{k}=-2t\cos(\pi k/k_0)$, one obtains the equation for wave function $\psi_{k}^{(+)}(n)$,
\begin{align}
\label{coe-equ}
-&t\left[\psi_{k}^{(+)}(n+1)+\psi_{k}^{(+)}(n-1)\right]+u^{\rm (II)}_{0,\xi}\psi_{k}^{(+)}(n)\delta_{n,0}+u^{\rm (II)}_{1,\xi}\psi_{k}^{(+)}(n)\left(\delta_{n,+1}+\delta_{n,-1}\right)=\epsilon_{k}\psi_{k}^{(+)}(n)
\end{align}
for $n=0,\pm1,\pm2,\cdots$. Furthermore, considering the expression of $h_\xi^{\rm (II)}$ and the out-going boundary condition, we can certainly express $\psi_{k}^{(+)}(n)$ as
\begin{eqnarray}
\label{boundary}
\psi_{k}^{(+)}(n)=
\left\{ \begin{aligned}
&e^{ik\lambda_{z}n/2}+\tilde{r}_{\xi}e^{-ik\lambda_{z}n/2},\hspace{0.2cm}n\leqslant-2\\
&\psi_{k}^{(+)}(0,\pm1),\hspace{0.2cm}n=0,\pm1\\
&\tilde{t}_{\xi}e^{ik\lambda_{z}n/2},\hspace{0.2cm}n\geqslant2
\end{aligned}
\right.,
\end{eqnarray}
where $\tilde{r}_{\xi}$ and $\tilde{t}_{\xi}$ are the reflection and transmission amplitude, respectively.
Substituting Eq.(\ref{boundary}) into Eq.(\ref{coe-equ}), one obtains the following equation set
\begin{eqnarray}
\label{eq-set}
\left\{ \begin{aligned}
&\epsilon_{k}\psi_{k}^{(+)}(0)=-t\left[\psi_{k}^{(+)}(1)+\psi_{k}^{(+)}(-1)\right]+u_{0,\xi}^{\rm (II)}\psi_{k}^{(+)}(0),\\
&\epsilon_{k}\psi_{k}^{(+)}(1)=-t\left[\tilde{t}_{\xi}e^{2ika}+\psi_{k}^{(+)}(0)\right]+u_{1,\xi}^{\rm (II)}\psi_{k}^{(+)}(1),\\
&\epsilon_{k}\psi_{k}^{(+)}(-1)=-t\left[\psi_{k}^{(+)}(0)+e^{-2ika}+\tilde{r}_{\xi}e^{2ika}\right]+u_{1,\xi}^{\rm (II)}\psi_{k}^{(+)}(-1),\\
&\epsilon_{k}\tilde{t}_{\xi}e^{2ika}=-t\left[\tilde{t}_{\xi}e^{3ika}+\psi_{k}^{(+)}(1)\right],\\
&\epsilon_{k}\left(e^{-2ika}+\tilde{r}_{\xi}e^{2ika}\right)=-t\left[\psi_{k}^{(+)}(-1)+e^{-3ika}+\tilde{r}_{\xi}e^{3ika}\right].
\end{aligned}
\right.
\end{eqnarray}
Upon solving Eq.(\ref{eq-set}), one find the reflection and transmission amplitudes $\tilde{r}_{\xi}$ and $\tilde{t}_{\xi}$. Similar to the 3D case, the even-wave and odd-wave scattering amplitudes $\tilde{f}_{\xi}^{\rm (e/o)}$ for this 1D tight-binding model are also given by 
\begin{align}
\tilde{f}_{\xi}^{\rm (e)}&=\frac{1}{2}\left(\tilde{r}_{\xi}+\tilde{t}_{\xi}-1\right)\label{f1e},\\
\tilde{f}_{\xi}^{\rm (o)}&=\frac{1}{2}\left(-\tilde{r}_{\xi}+\tilde{t}_{\xi}-1\right)\label{f1o}.
\end{align}
Furthermore, in the limit $k\rightarrow 0$, the amplitudes $\tilde{f}_{\xi}^{\rm (e/o)}$ also have the behavior
\begin{align}
\tilde{f}_\xi^{\rm (e)}(k)&=-\frac{1}{1+ik\tilde{a}_\xi^{\rm (e)}+{\cal O}(k^{3})};\label{aae}\\
\tilde{f}_\xi^{\rm (o)}(k)&=-\frac{ik}{ik+\frac{1}{\tilde{a}_\xi^{\rm (o)}}+{\cal O}(k^{2})}\label{aao},
\end{align}
where $\tilde{a}_\xi^{\rm (e/o)}$ are the even/odd scattering lengths for the 1D tight-binding model. 

Therefore, by solving Eq. (\ref{eq-set}) and using Eqs. (\ref{f1e}-\ref{aao}), we can derive $\tilde{a}_\xi^{\rm (e/o)}$, which are functions of $u_{0,\xi}^{\rm (II)},u_{1,\xi}^{\rm (II)}$ and $t$. 
By requiring $\tilde{a}_{\xi}^{\rm (e)}=a_{\xi}^{\rm (e)}$ and $\tilde{a}_{\xi}^{\rm (o)}=a_{\xi}^{\rm (o)}$, one finally obtains 
\begin{align}
\frac{u_{0,\xi}^{\rm (II)}}{t}&=-\frac{2\left(l_0^{2}-2l_0a_\xi^{\rm (o)}+a_\xi^{\rm (e)}a_\xi^{\rm (o)}\right)}{l_0\left(a_\xi^{\rm (e)}-a_\xi^{\rm (o)}\right)};\\
\frac{u_{1,\xi}^{\rm (II)}}{t}&=\frac{a_\xi^{\rm (o)}}{l_0-a_\xi^{\rm (o)}}
\end{align}
which are the Eq.(\ref{hubbardU0}) and (\ref{hubbardU1}) in the main text. 

Eq. (\ref{u0iii}) can be proved with the similar as above. Explicitly, this time we expect the 1D tight-binding model
\begin{eqnarray}
h_\xi^{\rm (I)}\equiv T_{\rm 1D}+ u_{0,\xi}^{\rm (I)}|0\rangle_g\langle 0|
\end{eqnarray}
($\xi=+,-$) can reproduce the even-wave scattering lengths $a^{\rm (e)}_{\xi}$ of the exact Hamiltonian $H$. Other calculations are same as the above ones for $h_\xi^{\rm (II)}$.
\end{widetext}

\bibliography{references}

\end{document}